\documentclass[showpacs, pra,onecolumn,preprintnumbers ,amsmath, amssymb, superscriptaddress, aps]{revtex4-2}
		\usepackage{color}
		\usepackage{amsmath,amssymb}
		\usepackage{pifont}
		\usepackage{amssymb}  
		\usepackage{bbold}
		\usepackage{float}
		\usepackage{subfloat}
		\usepackage[caption=false]{subfig}
		\usepackage{tikz}
		\usepackage{makecell}
		\usepackage{subfig}
		\usepackage{pifont}   
		\usepackage{graphicx} 
		\graphicspath{{Figures/}}
		\usepackage{dcolumn}  
		\usepackage{bm}       
		\usepackage{multirow} 
		\usepackage{placeins}
		\usepackage[colorlinks]{hyperref}
		\usepackage{mathtools}
		\usepackage{appendix}

		\def \be{\begin{align}}
			\def \ee{\end{align}}
		\def \bea{\begin{eqnarray}}
			\def \eea{\end{eqnarray}}
		
\begin{document}
			
			\title{Band structures and contact points  in phosphorene superlattice}
			\author{Jilali Seffadi}
			\affiliation{ Laboratory of Theoretical Physics, Faculty of Sciences, Choua\"ib Doukkali University, PO Box 20, 24000 El Jadida, Morocco}
			\author{Ilham Redouani}
			\affiliation{ Laboratory of Theoretical Physics, Faculty of Sciences, Choua\"ib Doukkali University, PO Box 20, 24000 El Jadida, Morocco}
			\author{Youness Zahidi}
			\affiliation{ MRI Labortory, Polydisciplinary Faculty, Sultan Moulay Selimane University, Khouribga, Morocco}
			\author{Ahmed Jellal}
			\affiliation{ Laboratory of Theoretical Physics, Faculty of Sciences, Choua\"ib Doukkali University, PO Box 20, 24000 El Jadida, Morocco}
			\affiliation{
				Canadian Quantum  Research Center,
				204-3002 32 Ave Vernon, BC V1T 2L7,  Canada}

			\begin{abstract}
				

We study the  band structures and the associated contact points for a phosphorene superlattice made up of two periodic areas. We use the boundary conditions to extract an equation describing the dispersion relation after obtaining the eigen-wavefunctions. We show that energy transforms into linear behavior near contact points, and fermions move at different speeds along $x$- and $y$- directions. It was discovered that the periodic potential caused additional Dirac points, which we located in $k$-space by establishing their positions. 
 We demonstrate that the barrier height and width can be used to adjust the energy gap and modify the contact points. 
 It might be that our findings will be useful in the development of phosphorene-based electronic devices. 
 
		\end{abstract}		

	\maketitle
	\section{INTRODUCTION}
	 
 The electrical characteristics of 2D materials like graphene have received enormous attention during the past few years. It has attracted great interest in the scientific community since its realization in the laboratory  \cite{ref1,ref2}. Such advantages result, especially, from the fact that a Hamiltonian of the Dirac type governs the physics of the low-energy  in graphene. Pure graphene has no gaps and may be represented by the massless Dirac equation. The Klein effect \cite{ref3,ref4,ref5} prevents electrons in graphene from being constrained by electrostatic potentials. Due to this characteristic, graphene cannot be used in electronic devices. The most challenging task is to figure out how to use electric fields to regulate the behavior of charge carriers in graphene. When a gap is produced between the conduction and valence bands, Klein tunneling is restrained, and the electronic properties of graphene change radically. To contain the massless Dirac particle in the graphene sheet, a gap in the spectrum's energy can be opened \cite{ref6,ref7,ref8}. Because of its straight bandgap, high mobility, and anisotropic transport and mechanical properties,  phosphorene can compensate well for the lack of performance of other 2D materials \cite{ref9,ref10}.

In 2014, two research groups published the first report on the isolation of phosphorene layers \cite{Brent2014,Guo2015}, and today, there are two basic techniques for making phosphorene: liquid exfoliation and scotch-tape cleavage \cite{Carvalho2016,Gusmao2017}.
	Unlike graphene, which has layers that are all plane, phosphorene is a weakly van der Waals-bonded layered material with puckered surfaces on each layer caused by the $sp^3$ hybridization of the $3s$ and $3p$ atomic orbitals \cite{ref11}. Phosphorene has an anisotropic puckered honeycomb band structure, which distinguishes it from that of graphene. Because of this, graphene has a unit cell with only two atoms, whereas there are four atoms in phosphorene \cite{ref12}. Additionally, phosphorene's strong anisotropy allows it to exhibit intriguing direction-dependent transport and optical features. Phosphorene's inherent band gap makes it potentially useful for adjusting the transition between conducting and insulating states. Due to its distinct electrical characteristics and possible uses in optoelectronic, nanoelectronic, and other applications,  phosphorene has received a lot of attention \cite{ref13,ref15}. In fact, a generation of anisotropic massless Dirac fermions and asymmetric Klein tunneling in few-layer  phosphorene superlattices was observed \cite{ref16}.
	It is shown that the highly anisotropic character of phosphorene in the presence of bias allows for a field-induced semiconductor-metal transition \cite{ref35}.
	
	
	Recent studies have used techniques including first-principles calculations, $k\cdot p$ methods, and tight-binding models to determine the electronic dispersion since it is crucial to have a complete understanding of the band structure and charge carrier dynamics in  phosphorene \cite{Rudenko2014}. These studies provided estimates of the energy gap for single-layer and multilayer phosphorene, as well as evidence of significant anisotropy in effective mass. 
	In this respect, we study the band structures of  phosphorene in the framework of a continuum model, derived as the long-wavelength limit of a newly proposed tight-binding mode. This study will be carried out by using a periodic potential, resulting in a phosphorene superlattice. We describe the conditions that allow this to occur and identify all Dirac points that appear when such potential exists. In addition, we numerically analyze our results by distinguishing between equal and unequal well and barrier widths under various conditions of the physical parameters to underline the basic features of our system. It turns out that the prediction that the periodic potential could be used to manipulate the system band structure has increased interest in phosphorene superlattices and opened up new avenues for the fabrication of phosphorene-based electronic devices.
	
%

 The present paper is organized as follows. In Sec.  \text{\ref{Sec2}}, we provide the theoretical model that underpins our approach and aims to derive the dispersion relation for the entire cell explicitly. In Sec. \text{\ref{DCPs}}, we apply the implicit function theorem to the dispersion relation  to find the contact points 
 and identify their locations. As an illustration, we consider the first Dirac point and derive the velocity components in the $x$- and $y$-directions, as well as show their behaviors against the barrier parameters. 
 In Sec. \ref{Sec4},  for the equal and unequal well and barrier widths, we present theoretical and numerical results based on various conditions. Our final objective is to compare these two situations and demonstrate that they are not equal. We conclude our results in Sec. \text{\ref{Secf}}.

\section{Theoretical model}\label{Sec2}

\begin{figure}[ht]
	\centering
	\includegraphics[width=14.4cm, height=4.6cm]{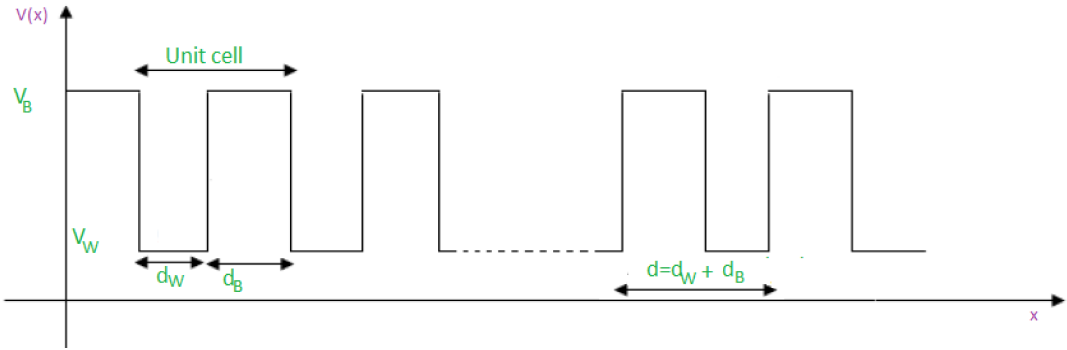}
	\caption{  (Color online) Schematic representation of the periodic potentials applied on the monolayer phosphorene. }\label{f01}
\end{figure}

We analyze the band structure of charge carriers in phosphorene by taking into account a periodic one-dimensional potential along the x-direction as depicted in Fig. \text{\ref{f01}}. We suggest that this structure can be created by giving the phosphorene layer a local top gate voltage \cite{ref20}. 
The one-dimensional superlattice potential structure characterized  by the  heights $(V_B, V_W)$ and widths $(d_B, d_W)$.
It is provided by the potential profile 

\begin{equation}
	\label{E1}
	V_j(x)=\left\{\begin{array}{llll}
		{V_B} & \mbox{if} &{id<x<id+d_B}\\
		{V_W} & \mbox{if} &  {id+d_B<x<(i+1)d}
	\end{array}\right.
\end{equation} 
where $ i \in \mathbb{N}$ and $\left(j=B,W \right)$. The period is defined as a series of wells and barriers that alternate in width $d$ $(d= d_B +d_W )$.

The Hamiltonian for monolayer phosphorene can be calculated using the low-energy five-hopping parameter tight-binding technique \cite{ref21, ref35, ref12, ref24}. This process yields
\begin{equation}\label{E2}
H(k)=
\begin{pmatrix}
	u_{0}+ \eta_{x} k_{x}^{2}+\eta_{y} k_{y}^{2} & \delta + \gamma_{x} k_{x}^{2}+\gamma_{y} k_{y}^{2}+i\chi k_x \\
	\delta+ \gamma_{x} k_{x}^{2}+\gamma_{y} k_{y}^{2}-i\chi k_x & u_{0}+ \eta_{x} k_{x}^{2}+\eta_{y} k_{y}^{2}\\
\end{pmatrix}%
\end{equation}
with the parameters quantities ${u}_0=-0.42$ {eV}, $\eta_{{x}}=0.58$ {eV} \AA$^2$, $\eta_{{y}}=1.01$ {eV}  \AA$^2$, $\delta=0.76$ {eV}, $\chi=5.25$ {eV} \AA, $\gamma_{{x}}=3.93$ {eV}  \AA$^2$ and $\gamma_{{y}}=3.83$ {eV} \AA.
%
The phosphorene superlattice is formed by n elementary cells positioned between the incidence and transmission areas. The Hamiltonian described below can be used to characterize our system in each $j$ region of the $n^{\text{th}}$ elementary cell
\begin{equation}\label{E3}
H_{j}(k)=H(k)+V_{j}(x)\mathbb{I}_2.
\end{equation}
By taking into account the conservation of the transverse wave vector $k_y$, we can express the eigenspinors of these quasi-particles traveling along the $\pm x$-directions in all areas as $\Phi_j(x, y) =\varphi_j(x)e^{ik_yy}$. 
%
By resolving the eigenvalue equation  $H_j\varphi_j=E_j\varphi_j$, the associated eigenspinors can be found as
\begin{align}
\label{E4}
\varphi_{j}=\Theta_j\cdot \Gamma_j
\end{align}
by setting the quantities
\begin{align}\label{E5}
\Theta_j=
\begin{pmatrix}
	e^{ik_{j}x}& e^{-ik_{j}x}\\
	z_{j}e^{ik_{j}x}& z_{j}^{-1}e^{-ik_{j}x}
\end{pmatrix}, \quad 
\Gamma_{j}=
\begin{pmatrix}
	{a_{j}} \\
	{b_{j}} \\
\end{pmatrix}
\end{align}
as well as 
\begin{align}
z_{j}=s_{j}e^{-i\theta_{j}}, \quad 	\theta_{j}=\arctan\left( \frac{\chi k_{j}}{\delta+\gamma_{y} k_{y}^{2}}\right)
\end{align}
where   $s_{j}={\mbox{sign}}{\left(E_j-V_j-u_{0}-\eta_{y}k_{y}^{2}\right)}$. 


We will then determine the probability that electrons will successfully pass through the  potential barriers. 
Specifically,  for the transmission and incidence regions,
we have
\begin{align}\label{E6}
\Gamma_{\sf in}=
\left(\begin{array}{c}
	{1} \\
	{r} 
\end{array}
\right), \quad \Gamma_{\sf tr}=
\left(\begin{array}{c}
	{t} \\
	{0} 
\end{array}
\right)
\end{align}
where $t$ and $r$ are the transmission and reflection coefficients, which can be determined by imposing   the continuity of the eigenspinors at interfaces. The transfer matrix formalism is the most convenient way to explain this process \cite{ref25}. As a result, we obtain 
\begin{align}\label{E7}
\begin{pmatrix}
	{1} \\
	{r} 
\end{pmatrix}
=\Lambda \cdot \begin{pmatrix}
	{t} \\
	{0} 
\end{pmatrix}
\end{align}
where the transfer matrix  $\Lambda$ is given by
\begin{equation}\label{E8}
\Lambda=\Theta_{\sf in}^{-1}[0]\cdot \lambda^{n} \cdot  \Theta_{\sf tr}[nd]=
\begin{pmatrix}
	\Lambda_{11}& \Lambda_{12}\\
\Lambda_{21}& \Lambda_{22}
\end{pmatrix}%
\end{equation}
and  $\lambda$ takes the form
\begin{align}\label{E9}
\lambda=
\Theta_B[0]\cdot \Theta_{B}^{-1}[d_{B}]\cdot \Theta_W[d_{B}]\cdot \Theta_{W}^{-1}[d].
\end{align}


Now,  using the Bloch theorem as well as the continuity at different interfaces to get the result
   \begin{align}\label{E10}
   	&
  \varphi_{B}(0)\Gamma_B= \varphi_{W}(d)\Gamma_W e^{ik_xd}
 \\
 &\label{E11}
   \varphi_{B}(d_B)\Gamma_B= \varphi_{W}(d_B)\Gamma_W
  \end{align}
  where $d$ is the length of the unit cell and $k_x$ is the Bloch wave vector. Thus, we can  express $\Gamma_W$ in terms of $\Gamma_B$ as 
   \begin{align}\label{E12}
 \Gamma_W &=\varphi_W^{-1}(d)\varphi_{B}(0)\Gamma_Be^{-i k_x d}\\
 &
  =\varphi_{W}^{-1}(d_B)\varphi_B(d_B)\Gamma_B \label{E13}.
  \end{align}
 Using  (\text{\ref{E5}}) along with (\text{\ref{E12}}) and (\text{\ref{E13}})
 to end up with the relation
    \begin{equation}\label{E14}
\Xi 
  \begin{pmatrix}
  {a_B} \\
  {b_B} \\
  \end{pmatrix}
 =0 
  \end{equation}
where the matrix $\Xi$ is provided by

\begin{equation}\label{E15}
\Xi=\varphi_W^{-1}(d)\varphi_{B}(0)e^{-ik_x d}-\varphi_{W}^{-1}(d_B)\varphi_B(d_B).
\end{equation}
Calculating $\det(\Xi)=0$  yields 
the dispersion relation
\begin{equation}\label{E16}
\cos(k_x d)=\cos(d_B k_B)\cos(d_W k_W)+\Upsilon\sin(d_B k_B)\sin(d_W k_W)
\end{equation}
such that the wave vectors and the parameter $\Upsilon$ are 
\begin{align}
&\label{E18}
k_B=\frac{1}{\chi}\sqrt{(E-V_B-\mu_0-\eta_y k^2_y)^2-(\delta+\gamma_y k^2_y)^2}\\
&\label{E18-1}
k_W=\frac{1}{\chi}\sqrt{(E-V_W-\mu_0-\eta_y k^2_y)^2-(\delta+\gamma_y k^2_y)^2}\\
	&\label{E17}
\Upsilon= \coth \theta_B \coth \theta_W - \frac{s_B s_W}{\sin{\theta_B}\sin{\theta_W}}.
\end{align}
In the analysis that will follow, a theoretical study will be done to shed light on the Dirac points. Also, we will see how the aforementioned findings can be applied to understanding band structures. In fact, a numerical analysis will be established to investigate the basic features of the present system.

\section{Dirac contact points}\label{DCPs}

To establish the Dirac contact points, we apply the implicit function theorem to the dispersion relation \eqref{E16}. 
Indeed, let us write (\text{\ref{E16}}) as an implicit function of the form
\begin{equation}\label{E-24}
	\begin{split}
		h(k_x,k_y,E)=0
	\end{split}
\end{equation}
which can be expanded  
near the contact point $(k_{xc},k_{yc},E_c)$ as follows
\begin{equation}\label{E25}
	\begin{split}
		h(k_x,k_y,E)=h(k_{xc},k_{yc},E_c)+\Delta P \nabla h(k_{xc},k_{yc},E_c)+\frac{1}{2}\Delta P^t\mathbb{H}h(k_{xc},k_{yc},E_c)\Delta P
	\end{split}
\end{equation}
and we have
\begin{equation}\label{E26}
	\begin{split}
		h(k_{xc},k_{yc},E_c)=0, \quad \nabla h(k_{xc},k_{yc},E_c)=0, \quad 	\Delta P=\begin{pmatrix}
			\Delta k_x \\
			\Delta k_y  \\
			\Delta E
		\end{pmatrix}
	\end{split}
\end{equation}
where $\mathbb{H}$ is the Hessian matrix of second-order partial derivatives of $h$, acting  as
\begin{equation}\label{E27}
	\begin{split}
		\mathbb{H}h(k_{xc},k_{yc},E_c)=\begin{pmatrix}
			\rho & 0 & 0 \\
			0 & \sigma & 0 \\
			0 & 0 & \tau
		\end{pmatrix}
	\end{split}
\end{equation}
and ($\rho$, $\sigma$, $\tau$) can be fixed  in terms of the physical parameters.
Now injecting  all into  \eqref{E25}  to end up with the interesting result 
\begin{equation}\label{E29}
	\begin{split}
		\rho \Delta k_x^2 + 	\sigma \Delta k_y^2+ 	\tau \Delta E^2=0 
	\end{split}
\end{equation}
or more explicitly 
\begin{equation}\label{E30}
	E-E_c=\pm \sqrt{-\frac{\rho}{\tau}(k_x-k_{xc})^2-\frac{\sigma}{\tau}(k_y-k_{yc})^2}
\end{equation}
which is the energy of a generic Dirac Hamiltonian with anisotropic
Fermi velocity
\begin{equation}\label{E32}
	\begin{split}
		v_y= \frac{1}{\hbar}\sqrt{-\frac{\sigma}{\tau}}, \quad 	v_x= \frac{1}{\hbar}\sqrt{-\frac{\rho}{\tau}}.
	\end{split}
\end{equation}
where $\hbar$ is  the  Planck constant.
This result is in agreement with that obtained in \cite{Soleimanikahnoj2017} by studying the tunable electronic properties of multilayer phosphorene and its nanoribbons.



As illustration, we consider the first contact point $(0,0,E_0)$, and therefore obtain
\begin{equation}\label{E33}
	\mathbb{H}h(0,0,E_0)=\begin{pmatrix}
		d^2 & 0 & 0 \\
		0 & \sigma_1 & 0 \\
		0 & 0 & \tau_1
	\end{pmatrix}
\end{equation}
such that 
$\sigma_1$ and $\tau_1$ are given by
\begin{align}
	&\label{E34}
	\sigma_1=4s_B s_W{\frac {\delta\sin \beta  \left( \epsilon \cos
			\beta  +\varsigma   \sqrt{V_B
				^{2}-4{\delta}^{2}} \sin  \beta \right) }{\chi \left( V_B^{2}-4{\delta}^{
				2} \right) ^{5/2}}}
	\\
	&\label{E35-1}
	\tau_1=-s_B s_W{\frac { \zeta \cos  \beta  \sin
			\beta  +	\sqrt{V_B^{2}-4{\delta}^{2}}\left(\xi 
		\cos^{2}  \beta +\upsilon \right)}{2{\chi}^{2} \left( V_B^{2}-4\delta^{2} \right) 
			^{7/2}}}	
\end{align}
and the quantities are
\begin{align}
	&\label{E35-3}
	\beta={\frac {d \sqrt{V_B^{2}-4{\delta}^{2}}}{4\chi}}
	\\
	&\label{E35-8}
	\epsilon =s_B s_W\left(V_B^{4}d\gamma_y-12 V_B^{2}d\delta^{2}\gamma_y+32 d{
		\delta}^{4}\gamma_y\right)+V_B^{4}d\gamma_y-4V_B^{2}d{\delta}^{2}\gamma_y	
	\\
	&\label{E35-7}
	\varsigma =4s_B s_W\gamma_yV_B^{2}\chi -4\gamma_y V_B^{2}\chi
	\\
	&\label{E36}
	\zeta =s_B s_W\left(-16V_B^{4}\chi d{\delta}^{2}+192 V_B^{2}\chi d{\delta}^{
		4}-512 \chi d{\delta}^{6}\right)-16 V_B^{4}\chi d{\delta}^{2}+64 V_B
	^{2}\chi d{\delta}^{4}
	\\
	&\label{E35-5}
	\xi=s_B s_W\left(64V_B^{2}{\chi}^{2}{\delta}^{2}+256{\chi}^{2}{\delta}^{4}\right)-
	192V_B^{2}{\chi}^{2}{\delta}^{2}+256{\chi}^{2}{\delta}^{4}
	\\
	&\label{E35-6}
	\upsilon= \left( s_B s_W-1 \right) \left(V_B^{6}{d}^{2}-4V_B^{4}{d}^{2}{\delta}
	^{2}  \right) -s_B s_W\left(64V_B^{2}{\chi}^{2}{\delta}^{2}-256{
		\chi}^{2}{\delta}^{4}\right)+192V_B^{2}{\chi}^{2}{\delta}^{2}-256{\chi}
	^{2}{\delta}^{4}	
\end{align}
Thus, we get the two velocity components 	 
\begin{equation}\label{E3-2}
	\begin{split}
		v_y= \frac{1}{\hbar}\sqrt{-\frac{\sigma_1}{\tau_1}}, \quad 	v_x=\frac{1}{\hbar} \sqrt{-\frac{d^2}{\tau_1}}.
	\end{split}
\end{equation}
\begin{figure}[H]
		\centering
\subfloat[$V_B=2, 2.5, 3$ eV]{
		\centering
		\includegraphics[scale=0.6]{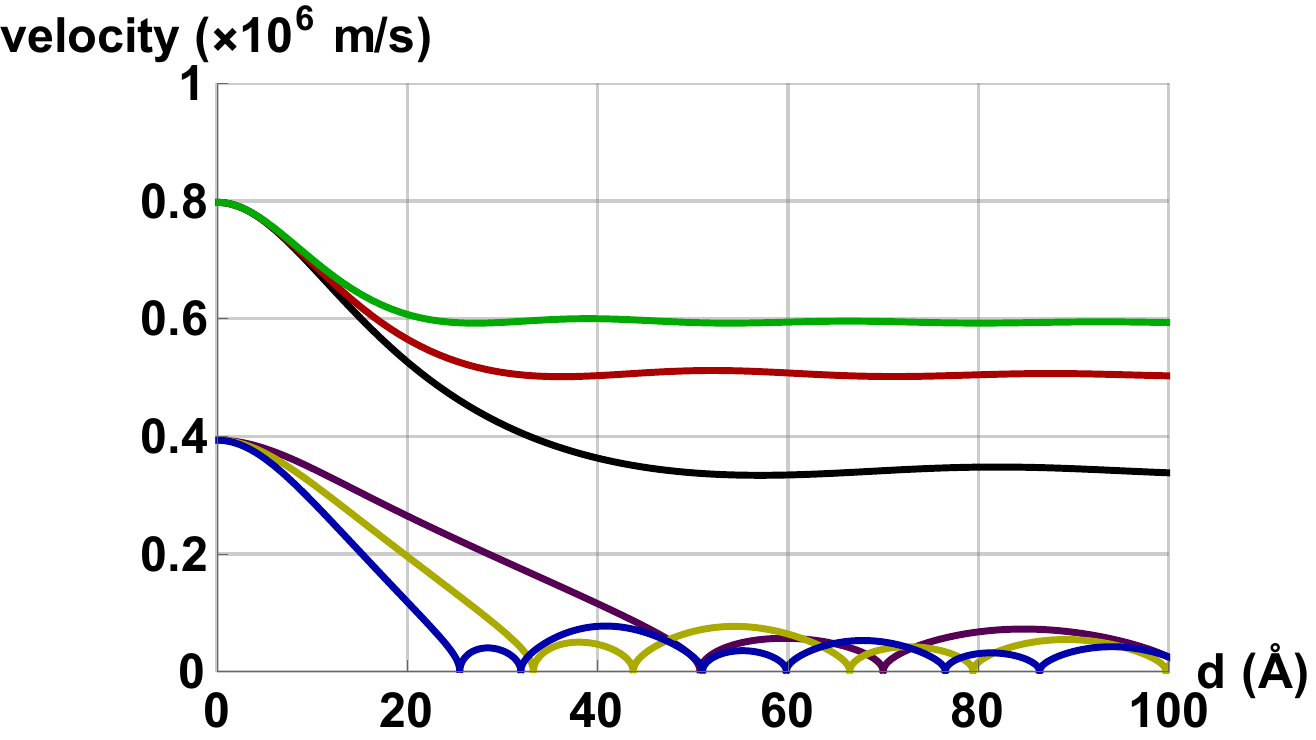}\label{v2-2}
	}
\subfloat[$d=2, 4, 6$ nm]{
		\centering
		\includegraphics[scale=0.6]{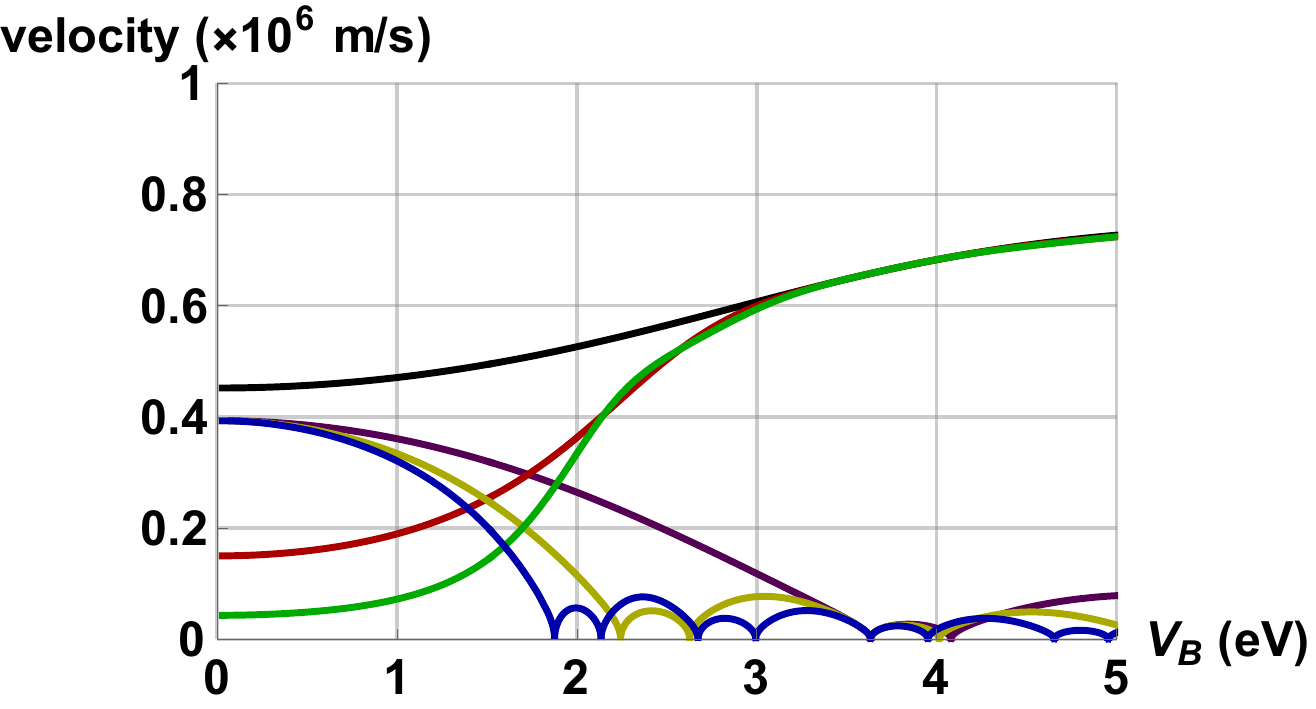}\label{v2-3}
	}
	\caption{ (color online) Velocity components ($v_x$, $v_y$)  versus barrier height  $V_B$ near the first Dirac point  for  $k_x=k_y=V_W=0$,  $E_0=\frac{V_B}{2}+\mu_{{0}}$ and $d_W=d_B$. 
		 (a): $V_B=2$ eV (black, purple), $V_B=2.5$ eV (red, yellow), $V_B=3$ eV (green, blue). (b): $d=2$ nm (black, purple), $d=4$ nm (red, yellow), $d=6$ nm (green, blue)
	}
	\label{velocity}
\end{figure}


To establish a deeper connection between the barrier parameters ($V_B$, $d$) and  the group velocities along the $k_x$-direction $v_x$ (black, red, green) and the $k_y$-direction $v_y$ (purple, yellow, blue), we plot    Fig. \ref{velocity} for the Dirac point $(0,0, E_0)$.
Fig. \ref{velocity}a  depicts the behavior of $v_x$ and $v_y$ versus the barrier width $d$ for three different values of its height $V_B$. 
It is clear that $v_x$ starts to decrease around $0.8\ 10^6 $ m/s, which is consistent with literature \cite{Chen2020, Soleimanikahnoj2017}. $v_x$ decreases as $d$ increases, eventually stabilizing at some fixed value dependent on $V_B$. 
On the other hand, as $d$ increases, $v_y$ reaches a maximum around $0.4\ 10^6$ m/s and then falls to zero. After a certain value of $d$, $v_y$ exhibits a variety of small oscillations that are strongly influenced by the barrier height $V_B$. 
The velocities are plotted against the barrier height $V_B$ for three different values of its width $d$
 in Fig. \ref{velocity}b.
 We can see that $v_x$ increases as $V_B$ increases, eventually reaching a constant value close to $0.8\ 10^6 $ m/s that is highly dependent on $d$. 
 In contrast, we see that $v_y$ behaves similarly to that shown in Fig. \ref{velocity}a, but with a slight variation in decreasing and oscillating as long as $V_B$ and $d$ increase. 
This is not surprising given that $v_x$ is strongly dependent on $d$ but $v_y$ is less so, according to \eqref {E3-2}. 
 Since the Fermi velocities control how rapidly electrons may flow between various components of nanoscale devices like transistors or memory cells, they have an impact on nanotechnology. 

\section{Analysis of band  Structures }\label{Sec4}

We will study the electrical band structures by distinguishing two cases. First, consider equal well and barrier widths, and then consider asymmetrical well and barrier widths. The valence and conductance mini-bands will be the main topics of study in the sections that follow. We will assume that the period of the phosphorene superlattice is constant $d=d_W+d_B$.
We will numerically investigate the dispersion relation for the phosphorene superlattice in \eqref{E16} in terms of the physical parameters $(V_B,d)$ describing the applied potential in order to explain the symmetrical impact of such a possibility. 
 
\subsection{Equal well and barrier widths $d_B=d_W$}\label{Sec3-1}


Before to numerically analyze the dispersion relation, let us investigate some interesting cases. Indeed, It is worth noting that when  $k_W=k_B$, we get 
\begin{equation}\label{E20}
	\begin{split}
		E_{{0}}=\frac{V_{{B}}}{2}+\mu_{{0}}+
		\eta_{{y}}k_{{y}}^{2}
	\end{split}
\end{equation}
which is nothing but the quadratic energy part of the pure phosphorene \cite{ref24} that will be established later on.
Now for  $k_x=0$ and  $d_W=d_B$, \eqref{E16}  reduces to 
\begin{equation}\label{E21}
	\begin{split}
		1=\cos^2(d_W k_W)+\Upsilon\sin^2(d_W k_W)
	\end{split}
\end{equation}
which is satisfied when  $k_W=\frac{n \pi}{d_W}$, and $n \in \mathbb{Z}$, resulting in quantized energy 
\begin{equation}\label{E-23}
	\begin{split}
		E_{n}=V_W+\mu_0+\eta_{y} k_y^2 \pm\sqrt{\left(\frac{\pi \chi}{d_W}\right)^2 n^2+\left(\delta +\gamma_y k_y^2\right)^2}
	\end{split}
\end{equation}
and the transverse momentum $ k_{{y}} $
\begin{align}\label{E22}
		k_{{yn }}= { \sqrt{ \frac{d_W\delta\,\gamma_{{y}}+d_W E_n\eta_{{y}}-d_W\eta_{{y}}
				\mu_{{0}}+ \sqrt{\left(4 \eta_\gamma \pi ^{2}{\chi}^{2}\right){n}^{2}  +2\,d_W^{2}\delta\,\eta_{{y}}\gamma_{{y}}
					\left( E_n-\mu_{{0}} \right) +\kappa_\mu+d_W^{2}\kappa_\eta 
				} }{{d_W\eta_\gamma }} }}
\end{align}
where we have set  $\eta_\gamma=\eta_{y}^{2}-\gamma_{y}^{2}$, $\kappa_\eta= 
	\eta_{y}^{2}{\delta}^{2}+\gamma_{{y}}^{2}{\mu_{{0}}}^{2}  $, and $\kappa_\mu=E_n d_W^{2} \gamma_{{y}}^2
\left( E_n-2\,\mu_{{0}} \right)  ^{2}$. 
One obtains from the zeros of \eqref{E22}
\begin{equation}\label{E23}
	\begin{split}
		V_{B}=V_{Bn}=2\sqrt{\left(\frac{2 \pi \chi}{d}\right)^2 n^2 +\delta^2}
	\end{split}
\end{equation}
which are the $V_B$ values that fill the minigap when $k_y=0$. It is worth noting that $V_B$ has a critical value, which is  $V_{Bc}=V_{B1}$. We make the positions of the contact points  obvious by noting that when $V_B$ exceeds $V_{Bc}$, two additional Dirac points appear at $k_y \neq 0$. 
The energy minigap at $k_y = 0$ closes at  $V_{B}=V_{B2}$, creating three contact points in the system. When $V_B>V_{B2}$, the minigap at $k_y = 0$ opens, but two additional Dirac points appear instead, leaving the system with four contact points, and so on. Then, in the generic case, $E_n$ and $k_{yn}$ indicate the location of the contact point that is closest to $k_y = 0$ for $V_B=V_{Bn}$ or $V_{Bn}< V_B <V_{Bn+1}$. As a result, the contact points are located as 
\begin{align}
(E,k_x,k_y) = (E_n, 0,k_{yn} ), (E_{n-1}, 0,k_{yn-1}), \cdots , (E_1, 0,k_{y1} ).	
\end{align}

	Now, by requiring  $d_B=0$,  we may rewrite  \eqref{E16} as follows: 
	\begin{equation}\label{E-24-2}
		\begin{split}
			\cos(d_W k_W)-	\cos(k_x d_w)=0.
		\end{split}
	\end{equation}
	which produces the energy of phosphorene as found in \cite{ref24}
	\begin{equation}\label{E-24-4}
		\begin{split}
			E_{\text{phos}}=\mu_0+\eta_y k^2_y+V_W\pm \sqrt{k^2_x  \chi^2+(\delta+\gamma_y k^2_y)^2}.
		\end{split}
	\end{equation}
	For the Dirac points $\left(k_x=\frac{m\pi}{d_W},k_y=0\right)$, \eqref{E-24-2} gives 
	\begin{equation}\label{E-24-5}
		\begin{split}
			\cos\left( \frac{d_W}{\chi}\sqrt{(E-\mu_0)^2-\delta^2}\right)=	\cos(m\pi), \quad m\in \mathbb{Z}
		\end{split}
	\end{equation}
	showing the quantized energy 
	\begin{equation}\label{E-pristine}
		\begin{split}
			E_m= \mu_0\pm \sqrt{\left({\frac {\pi{\chi}}{{d_W}}}\right)^2{m}^{2}+{\delta} 
				^{2}}.
		\end{split}
	\end{equation}

\begin{figure}[H]
	\centering
	\subfloat[$d=9.5$ nm]{
	\centering
	\includegraphics[scale=0.42]{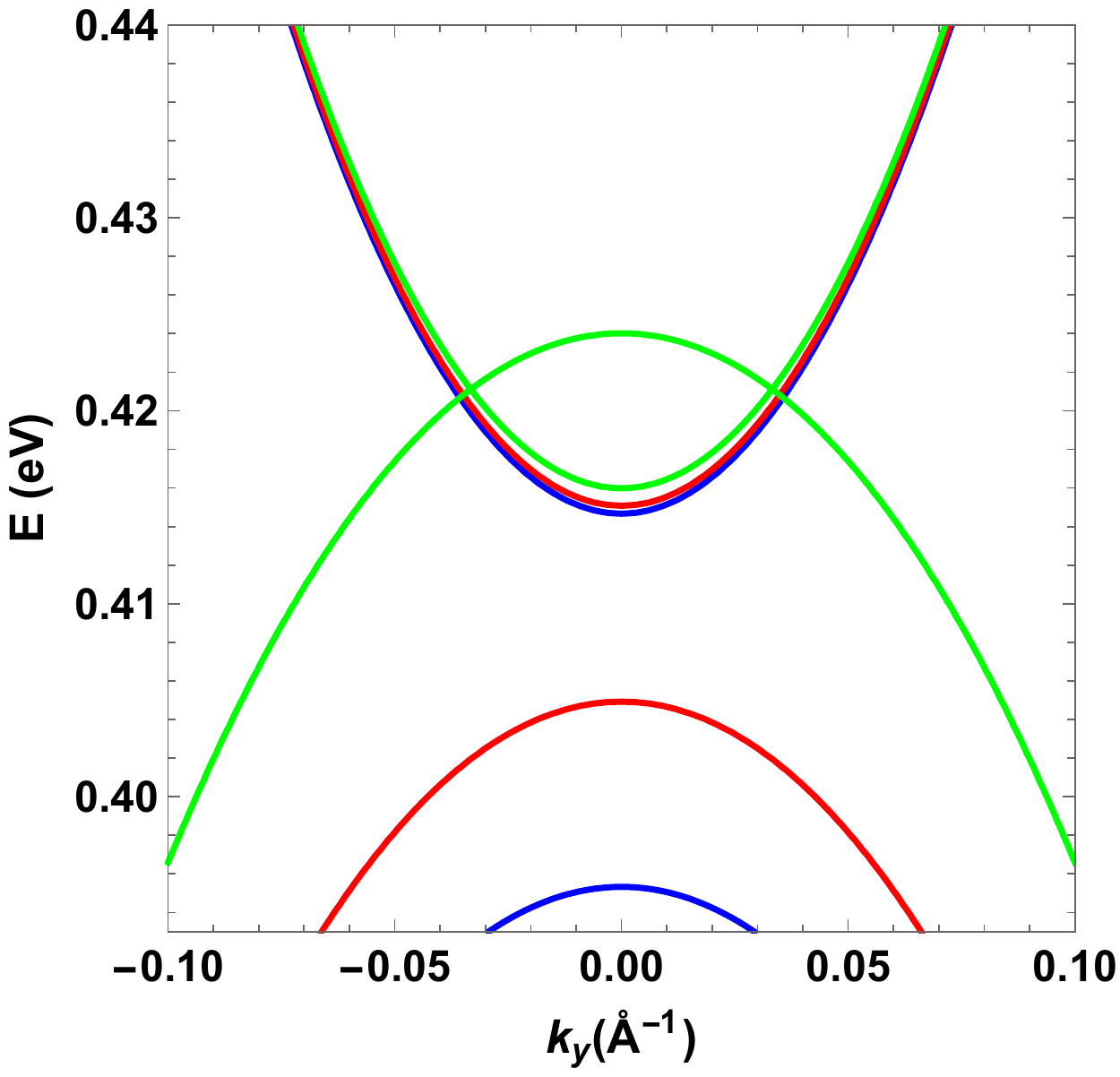}\label{f2-1}
}\subfloat[$d=10$ nm]{
\centering
\includegraphics[scale=0.42]{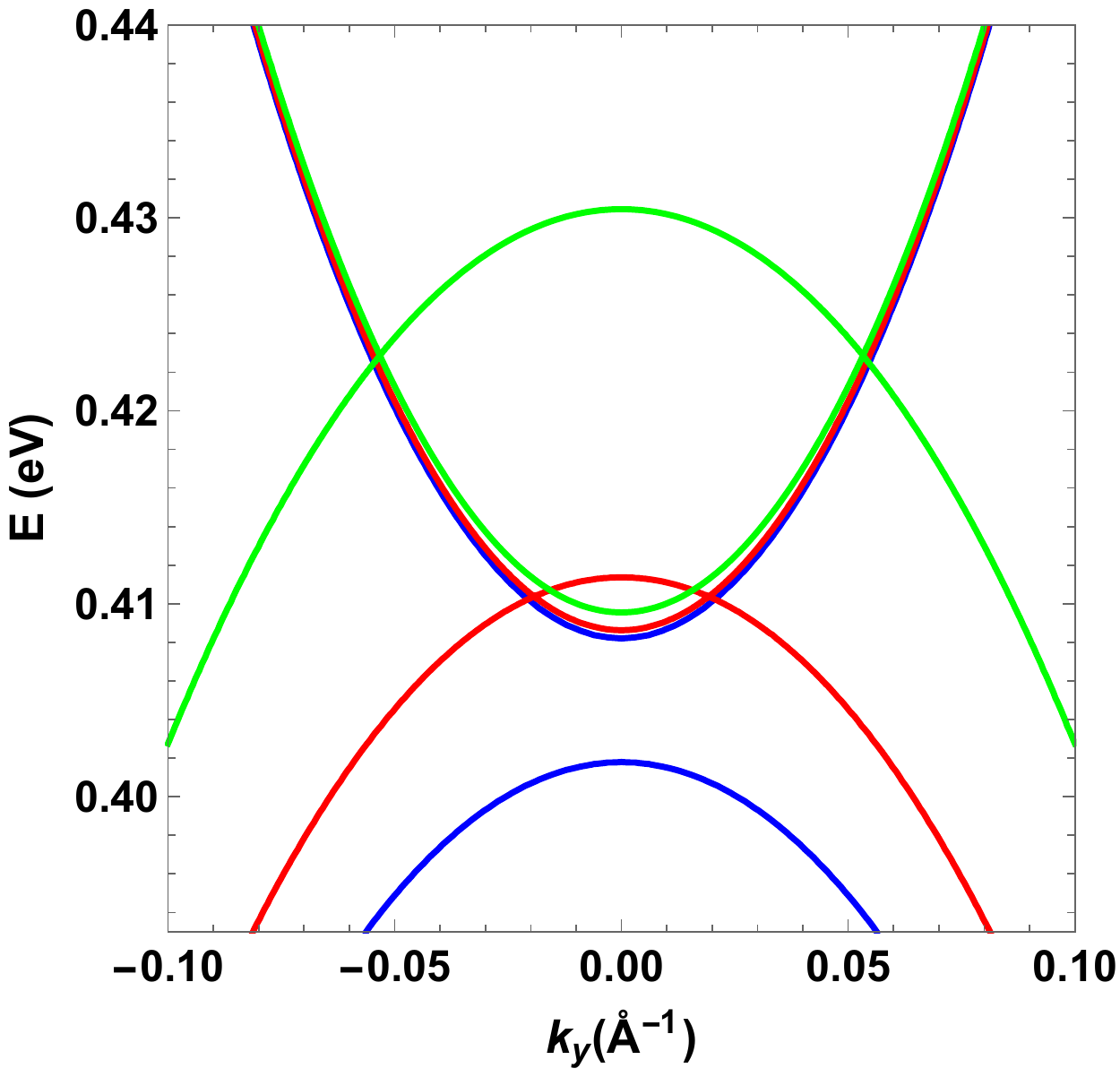}\label{f2-2}
}\subfloat[$d=10.5$ nm]{
\centering
\includegraphics[scale=0.42]{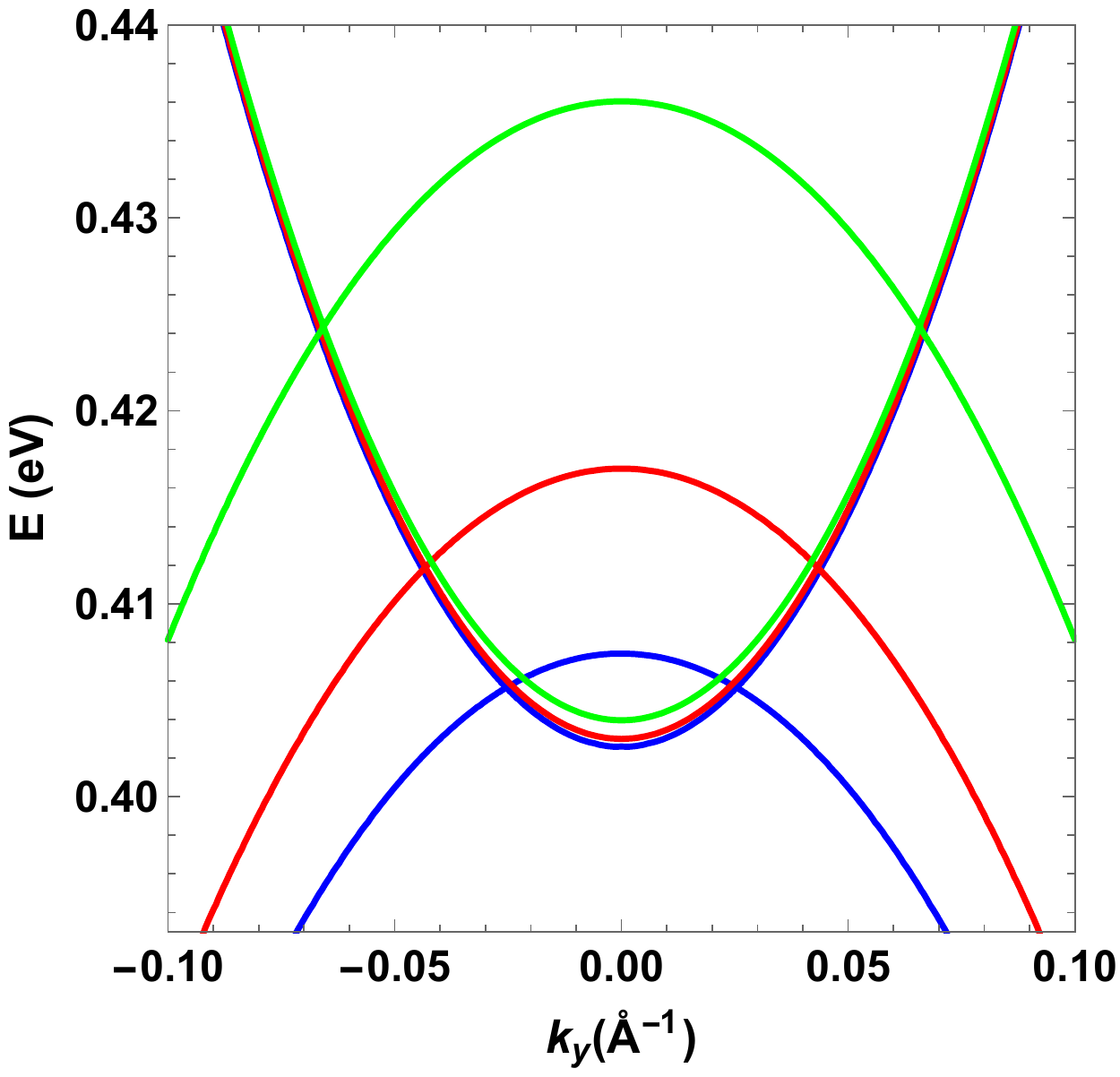}\label{f2-3}
}

	\caption{(color online) The dispersion relation \eqref{E16} as a function of $k_y$ for $k_x=0$, $V_B=1.65$ eV (blue), $V_B=1.66$ eV (red), $V_B=1.68$ eV (green) with (a): $d=9.5$ nm, (b):   $d=10$ nm, 
		(c):   $d=10.5$ nm.
	}\label{f2}
\end{figure}

The energies of holes and electrons are plotted as a function of the transverse momentum $ k_y $ with $ k_x = 0 $ and $ d_B = d_W = 4.75 $ nm for $ V_B = 1.65 $ eV (blue), 1.66 eV (red), and 1.68 eV (green) in Fig. \ref{f2}a. We see it is attainable to downgrade and control the gap until it is gapless by elevating the height potential. When $V_B$ reaches a critical value $V_{Bc}=1.68$ eV, we notice two Dirac points within the band structure, one on the positive side and the other on the negative side. {Same results are obtained in studying  the generation of anisotropic massless Dirac fermions and asymmetric Klein tunneling in few-layer black phosphorus superlattices \cite{ref16}}. We can also see, that the Dirac points move up in energy as $V_B$ increases. We show that the energy is symmetric for $k_y=0$. We can clearly see that the critical value becomes $V_{Bc}=1.66$ eV and then $V_{Bc}=1.65$ eV in Figs. \ref{f2}(b,c), where $ d = 10 $ and 10.5 nm, respectively, instead of $ d = 9.5 $ nm in Fig. \ref{f2}a. The same results were attained for the electronic structure of a graphene superlattice with massive Dirac fermions \cite{ref26}. This is in contrast to the electronic structure of a graphene superlattice with a modulated Fermi velocity \cite{ref27}. As we increase $V_B$ above $V_{Bc}$, a band gap appears at $k_y=0$ and grows monotonically with increasing $V_B$, as seen in \cite{ref28}.

\begin{figure}[ht]
	\centering
	\subfloat[$V_B=1.642$  eV]{
		\centering
		\includegraphics[scale=0.37]{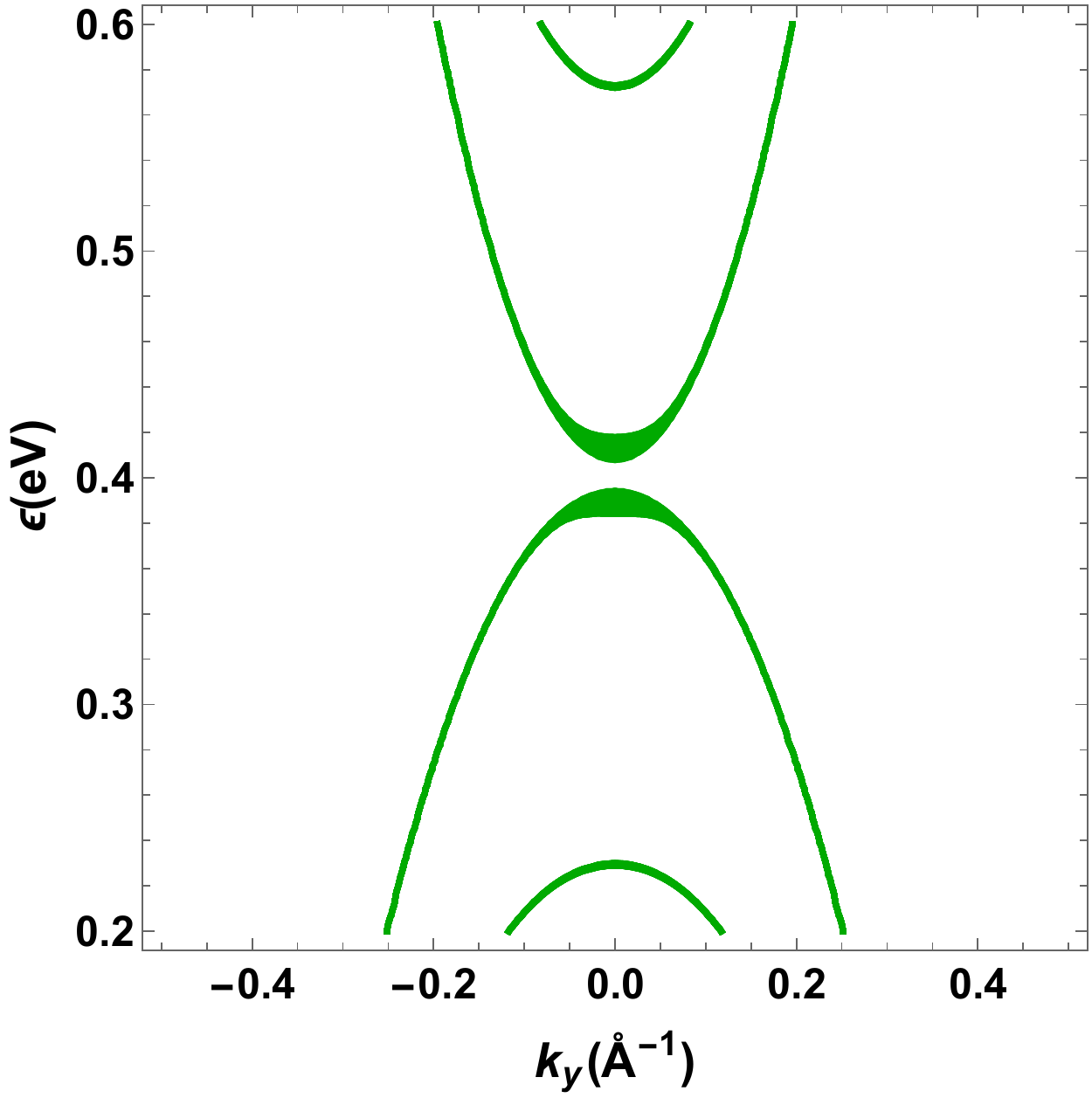}\label{f3-1}
	}\subfloat[$V_B=1.68$  eV]{
		\centering
		\includegraphics[scale=0.37]{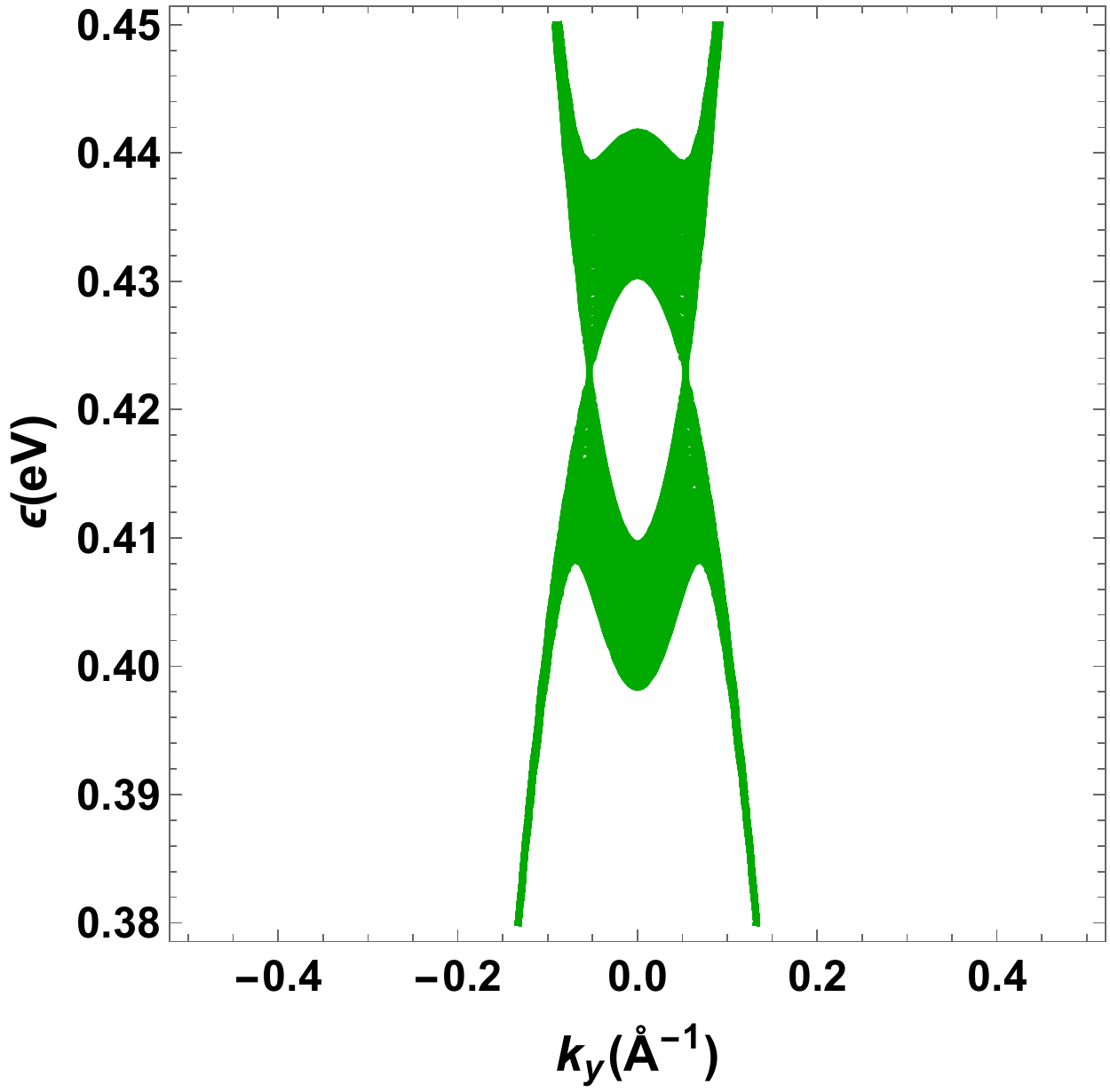}\label{f3-2}
	}\subfloat[$V_B=3$ eV]{
	\centering
	\includegraphics[scale=0.37]{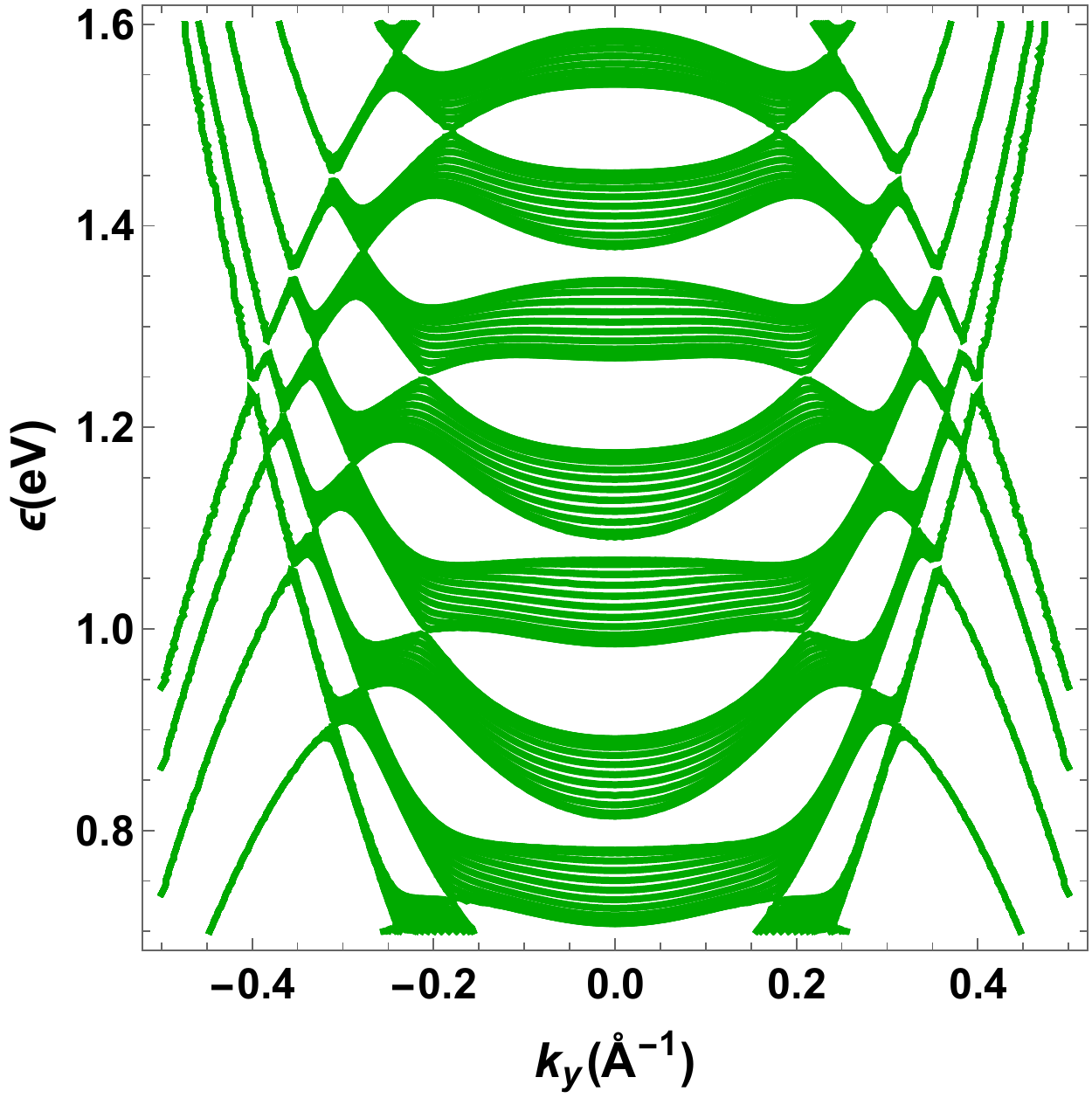}\label{f3-3}
}

	\caption{(color online) The dispersion relation \eqref{E16} as a function of $k_y$ for $k_x=V_W=0$, $d_B=d_W=\frac{d}{2}=5$ nm with (a):  $V_B=1.642$ eV, (b): $V_B= 1.68$ eV, (c):   $V_B=3$ eV.
	}\label{f3}
\end{figure}

Fig. \ref{f3}  demonstrates the absence of an original Dirac point and any vertical Dirac points in our study, while we have a gap.  However, it is evident that the potential has an impact on how Dirac points occur. It is clearly seen that the number of Dirac points increases as $V_B$ increases. Fig. \ref{f3}a shows that our system behaves like pure phosphorene, with a gap for a tiny barrier $V_B$ ($V_B<V_{Bc}=1.66$ eV) and energy that is quadratic along the $k_y$-direction. Fig. \ref{f3}b shows the development of two Dirac points when $V_B$ reaches the critical value $V_{Bc}$, and Fig. \ref{f3}c supports our findings by showing that when $V_B$ grows, the number of Dirac points similarly increases. The band structure at Dirac points changed from seeming parabolic to linear, as seen in Figs. \ref{f3}(b,c).


\begin{figure}[H]
	\centering
	\subfloat[$V_B=1.68$ eV, $d=10$ nm]{
		\centering
		\includegraphics[scale=0.5]{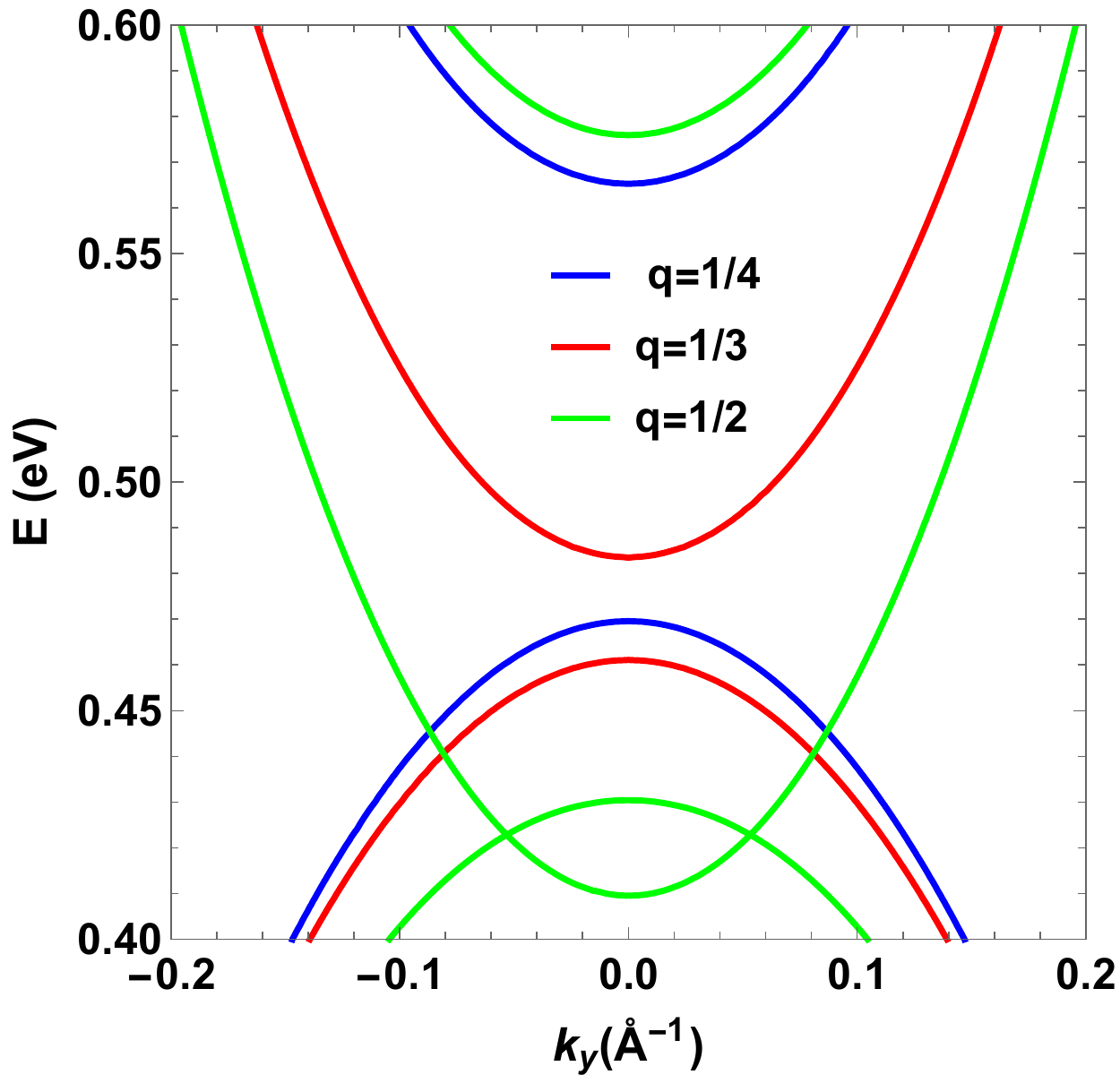}
	}
	\caption{The dispersion relation (\text{\ref{E16}}) as a function of $k_y$  for $d_W=q d$, $d_B=(1-q)d$   $d=10$ nm, and	 $k_x=0$.
	}\label{f4}
\end{figure}

To emphasize the importance of barrier and well widths, Fig. \ref{f4} depicts the effect of distance $q$ for $V_B = 1.68$ eV and $d = 10$ nm. We observe that by increasing $q$, it is possible to downgrade and control the gap until the system becomes gapless. Furthermore, the minigap at $k_y = 0$ opens for $q = \frac{1}{2}$ (green). 
 However, at various values of $k_y$ and for the parameters selected here, we notice there are extra Dirac points. As $q$ rises, the gap narrows until Dirac points are reached. 
 We conclude that there are two extreme cases to consider. Indeed, the system behaves like a pristine phosphorene with a direct band gap for $q< \frac{2}{5}$, whereas $q\geq \frac{2}{5}$ exhibits extra Dirac points.

%
%

\begin{figure}[H]
	\centering
	\subfloat[$d=10$ nm]{
		\centering
		\includegraphics[scale=0.5]{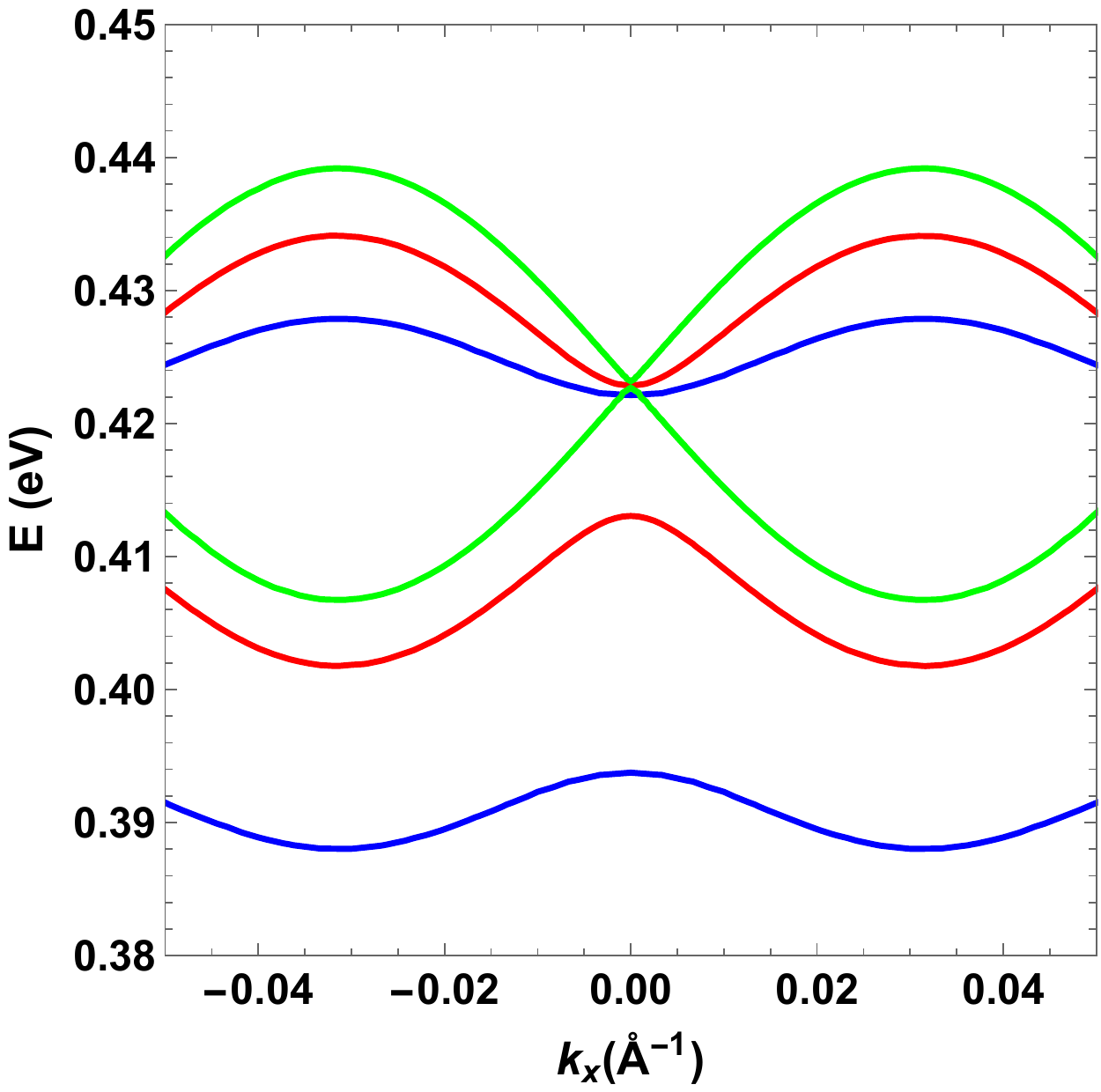}\label{f5-1}
	}\subfloat[$d=10$ nm]{
		\centering
		\includegraphics[scale=0.5]{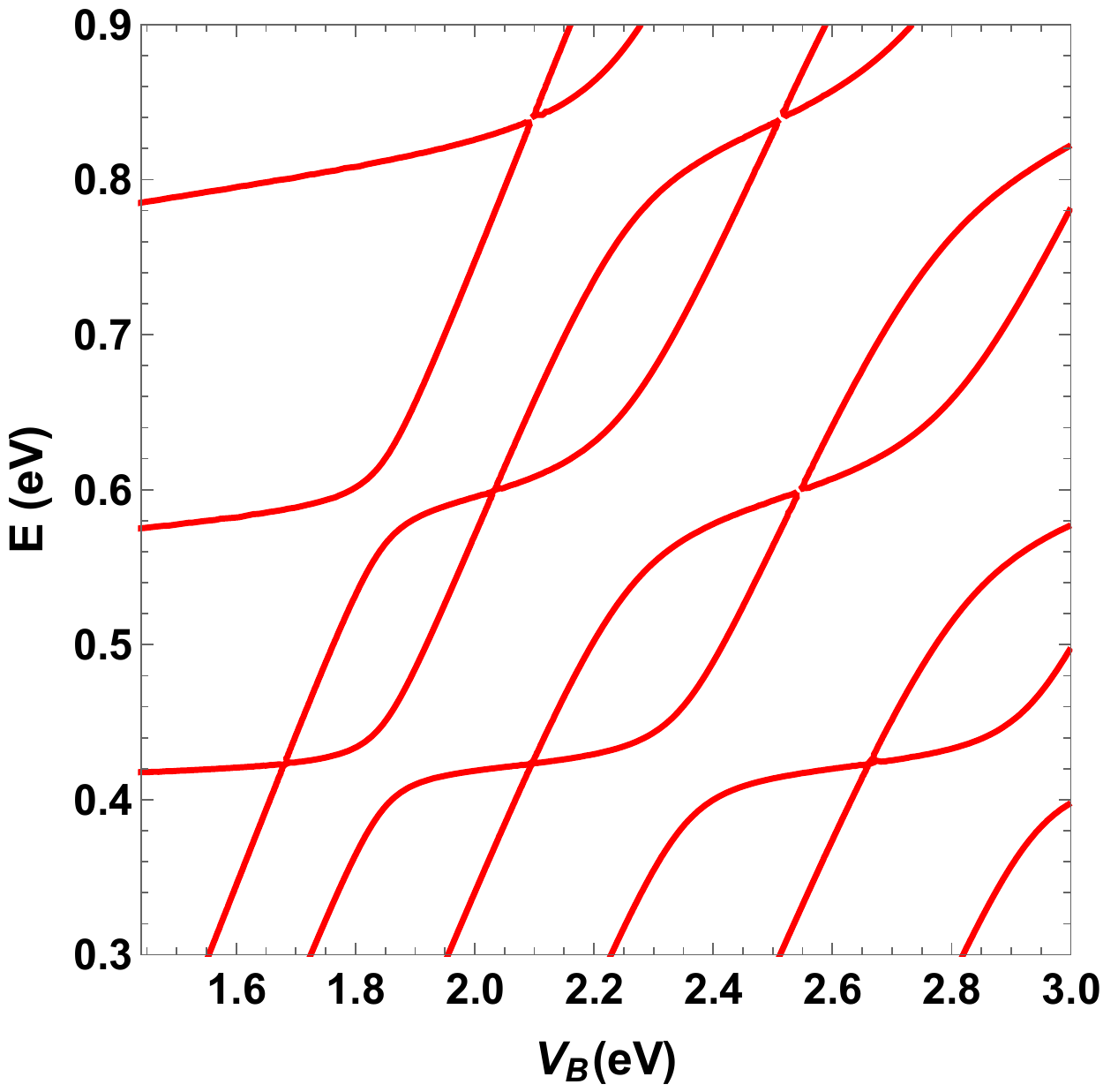}\label{f5-2}
	}
		\caption{(color online) (a): The dispersion relation (\text{\ref{E16}}) as a function of $k_x$ for  $k_y=0.054 \ \AA$, $V_B=1.65$ eV (blue), $V_B=1.67$ eV (red), $V_B=1.68$ eV  (green) and  $d=10$ nm. (b): The dispersion relation (\text{\ref{E16}}) as a function of $V_B$ for  $d=10$ nm, 	 $k_x=0$, and	$k_y=0.054 \ \AA^{-1}$.
	}\label{f5}
\end{figure}

In Fig. \ref{f5}a the plots of dispersion relation is  shown as a function of $k_x$ for $k_y=0.054 \ \AA$, $V_B=1.65$ eV (blue), $V_B=1.67$ eV (red), $V_B=1.68$ eV  (green) and  $d=10$ nm. The gap can be adjusted until it disappears by varying $V_B$, as seen when $V_B=1.68$ eV (green). The electron and hole energies shift upward with increasing potential, as can be seen. Nevertheless, The electron energy shift is not the same as the hole energy shift, which means that for different values of $V_B$ there are different gaps, some of which may be zero. Noteworthy is the fact that the Dirac point is obtained at $k_x = 0$.
The electron and hole energies are plotted as a function of $V_B$ in Fig. \ref{f5}b for $k_x =V_W=0$, $k_y=0.054\ \AA^{-1}$, and $d_B=d_W=5$ nm. It is feasible to see that the energy oscillates and exhibits gapless behavior at specific values of $V_B$, which depend on different parameters. It is critical to remember that these findings are similar to those reported in \cite{ref19}. 

\begin{figure}[H]
	\centering
	\subfloat[$d=10$ nm, $V_B=1.68$ eV, $k_x=V_W=0$]{
		\centering
		\includegraphics[scale=0.45]{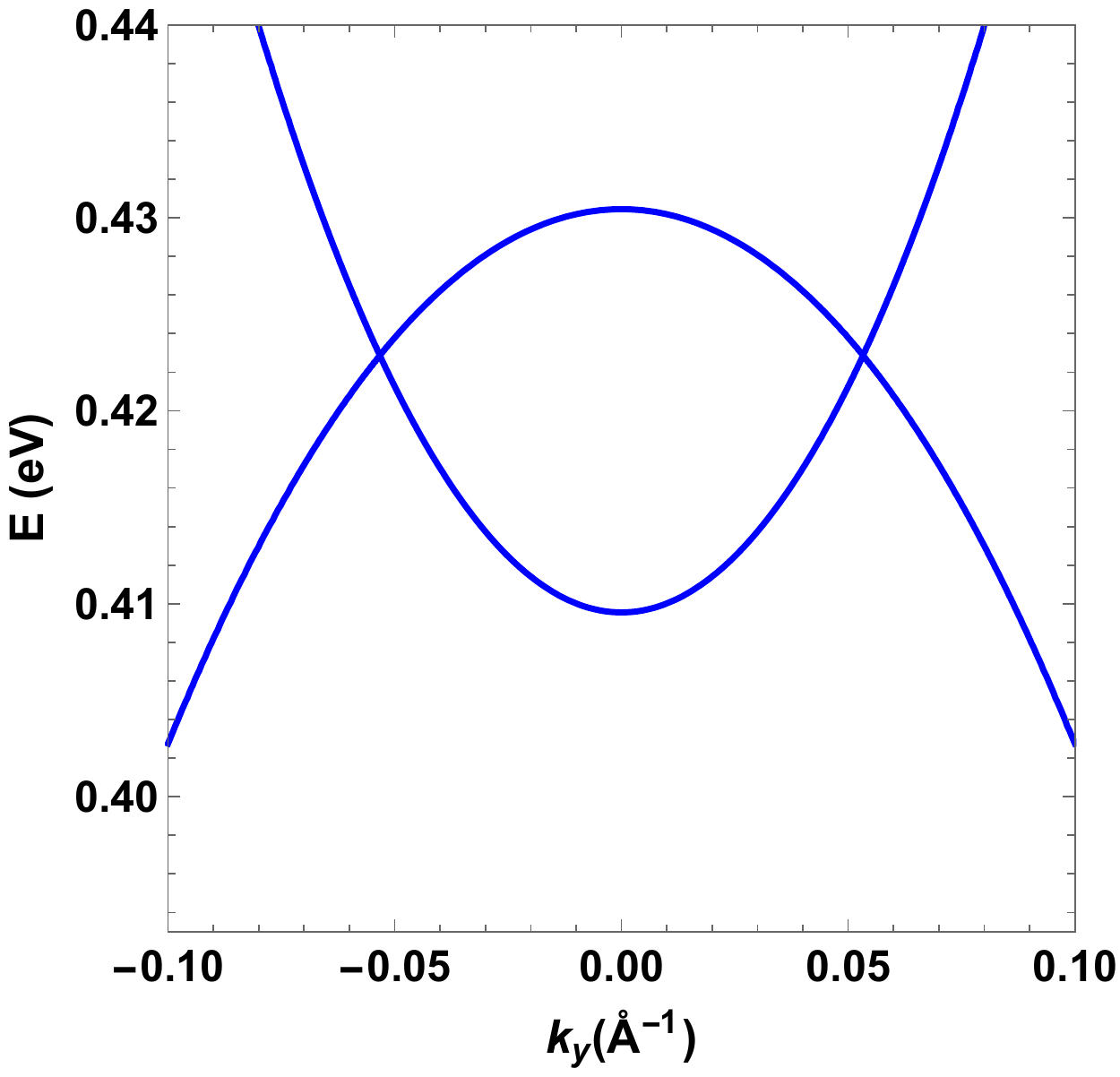}\label{f6-1}
	}\subfloat[$d=10$ nm, $V_B=1.68$ eV]{
		\centering
		\includegraphics[scale=0.45]{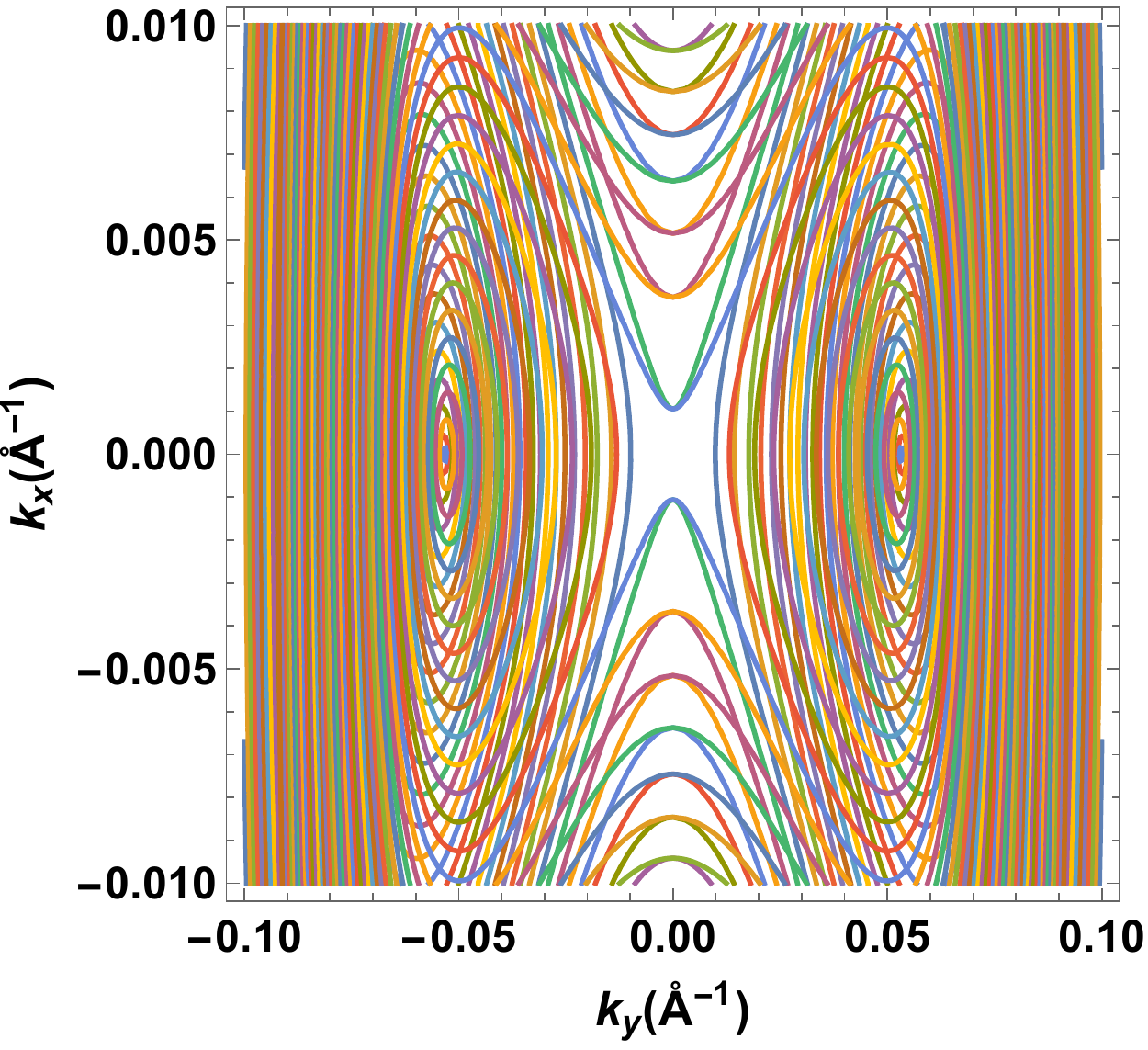}\label{f6-2}
	}
	\subfloat[$d=10$ nm, $V_B=1.68$ eV]{
				\centering
				\includegraphics[scale=0.45]{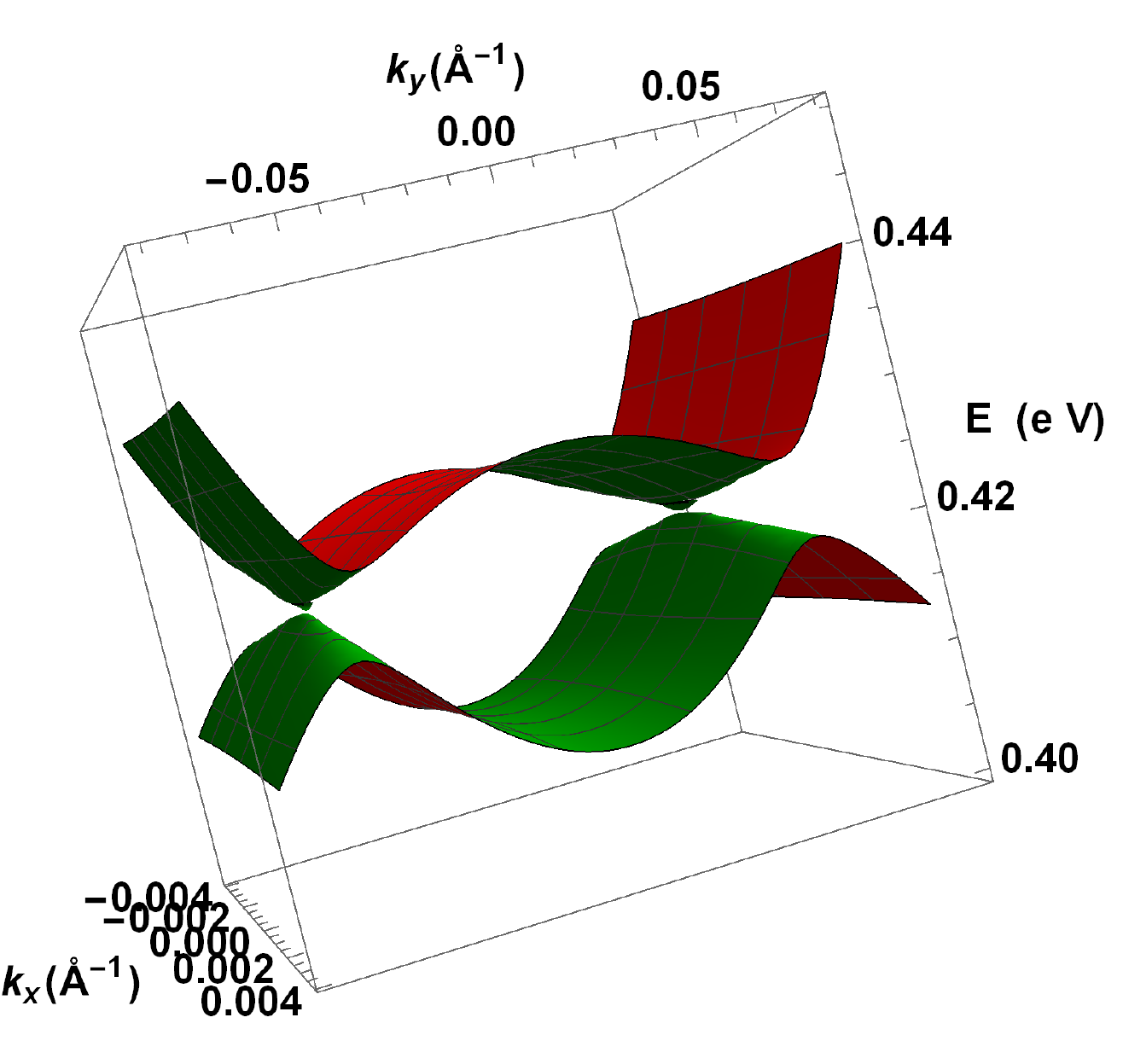}\label{f6-3}
		}

	\caption{(color online) (a): The dispersion relation (\text{\ref{E16}}) as a function of $k_y$ for $k_x=V_W=0$ and  $V_B=1.68$ eV. (b): Contour plot of the energy band $E \in[0.38,0.45] $ eV with step $5.10^{-4}$, $d=10$ nm and $V_B=1.68$ eV.  (c): Contour plot 3D of the energy band as a  function of $k_y$ and $k_x$.
	}\label{f6}
\end{figure}

The band structure around the Dirac points is depicted in Fig. \ref{f6}a for $d=10$ nm and $V_B=1.68$ eV. There are  two Dirac points, which they are isotropic and symmetrically located with respect to the point $k_y=0$.  The projection of the low-energy band centered on the Dirac points with $V_B = 1.68$ eV is shown in  Fig. \ref{f6}b. We can have two Dirac points by selecting such a value for $V_B$. The new Dirac cones have more clearly defined isotropic behavior. Energy is symmetric for $k_y=0$ as shown more clearly by the valence and conduction bands of the spectrum in Fig. \ref{f6}c. Moreover, at normal incidence there is no original Dirac point (ODP), which means that there is a gap in contrast to the case for \cite{ref19} where there is an ODP in pristine graphene. The most affecting characteristic observed here is that instead of having an  original zero-energy Dirac point (DP) of the pristine graphene at $k_y = 0$ there is the appearance of a pair of new contact points. This effect can also be seen at contact points in the energy band structure of bilayer graphene superlattices \cite{ref31} and graphene superlattices with  periodically modulated Dirac gaps \cite{ref32}. 

\subsection{Non equal well and barrier widths $d_B\neq d_W$ }\label{Sec3-2}


For the case $d_B \ne d_W$, we use also  \eqref{E21} to end up with 
\begin{align}
	&\label{E37}
	k_W =\frac{1}{\chi}\sqrt{(E-V_W-\mu_0-\eta_y k^2_y)^2-(\delta+\gamma_y k^2_y)^2} =\frac{m\pi}{d_W}\\
	&\label{E38}
	k_B =\frac{1}{\chi}\sqrt{(E-V_B-\mu_0-\eta_y k^2_y)^2-(\delta+\gamma_y k^2_y)^2} =\frac{m\pi}{d_B}
\end{align}
where $m\in \mathbb{Z}$. We get by subtracting  (\text{\ref{E38}}) from (\text{\ref{E37}}) the quantized energy
\begin{equation}\label{E39}
	E_m={\frac {{\pi }^{2}{\chi}^{2}{m}^{2}\xi_{-} +2V_{{B}}d_{{B}}^{2}  d_{{W}}^{2}\epsilon_0\mp \sqrt{{\pi }^{2}{\chi}^{2} 
				{\eta_{{y}}}^{2} \left[ {\pi }^{2}{\chi}^{2} \xi_{-}^{2}{m}^{2}-V_{{B}}^{2}d_{{B}}^{2}  d_{{W}}^2
				\left( 2\xi_{+}-V_{{B
				}}^{2}d_{{B}}^{2}d_{{W}}^{2}\eta_{{y}}^{2} \right) \right]}}{2\gamma_{{y}}d_{{W}}^{2}d_{{B}}^{2}V_{{B}}}}
\end{equation}
and we have set the quantities $\xi_{\pm}=d_{B}^{2}\pm d_{W}^{2}$ and $\epsilon_0= -
\delta\,\eta_{{y}}+\gamma_{{y}}\mu_{{0}}+\frac{V_{{B}}\gamma_{{y}}}{2}$.
Injecting (\text{\ref{E39}}) into (\text{\ref{E37}}) to find
\begin{equation}\label{E40}
	k_{{y}}=k_{ym}={\frac { \sqrt{d_{{W}}\eta_\gamma  \left( d_{{W}} \left( \delta\,\gamma_{{y}}+E_m\eta_{{
						y}}-\eta_{{y}}\mu_{{0}} \right) + \sqrt{{\pi }^{2}{\chi}^{2}{m}^{2}
					\eta_\gamma + \kappa ^{2}-\gamma_{{y}}
						d_{{W}}^{2}\mu_{{0}} \left( 2\,\delta\,\eta_{{y}}+2\,E_m\gamma_{{
							y}}-\gamma_{{y}}\mu_{{0}} \right) } \right) }}{d_{{W}} \eta_\gamma }}
\end{equation}
where $\eta_\gamma=\ \eta_{y}^{2}-\gamma_{y}^
{2}$ and $\kappa= \delta\,d
_{{W}}\eta_{{y}}+E_m d_{{W}}\gamma_{{y}}$.
From the zeros of (\text{\ref{E40}}), we get
\begin{equation}\label{E41}
	V_{{{\it B}}}= V_{{{\it Bm}}}=\sqrt{{\left(\frac {\pi \chi}{d_B}\right)^2{m}^{2}}+{\delta}^{2}}- \sqrt{{\left(\frac {\pi \chi}{d_W}\right)^2{m}^{2}}+{\delta}^{2}}
\end{equation}

\begin{figure}[H]
	\centering
	\subfloat[$d=9 $ nm, $d_W=2 d_B$]{
		\centering
		\includegraphics[scale=0.55]{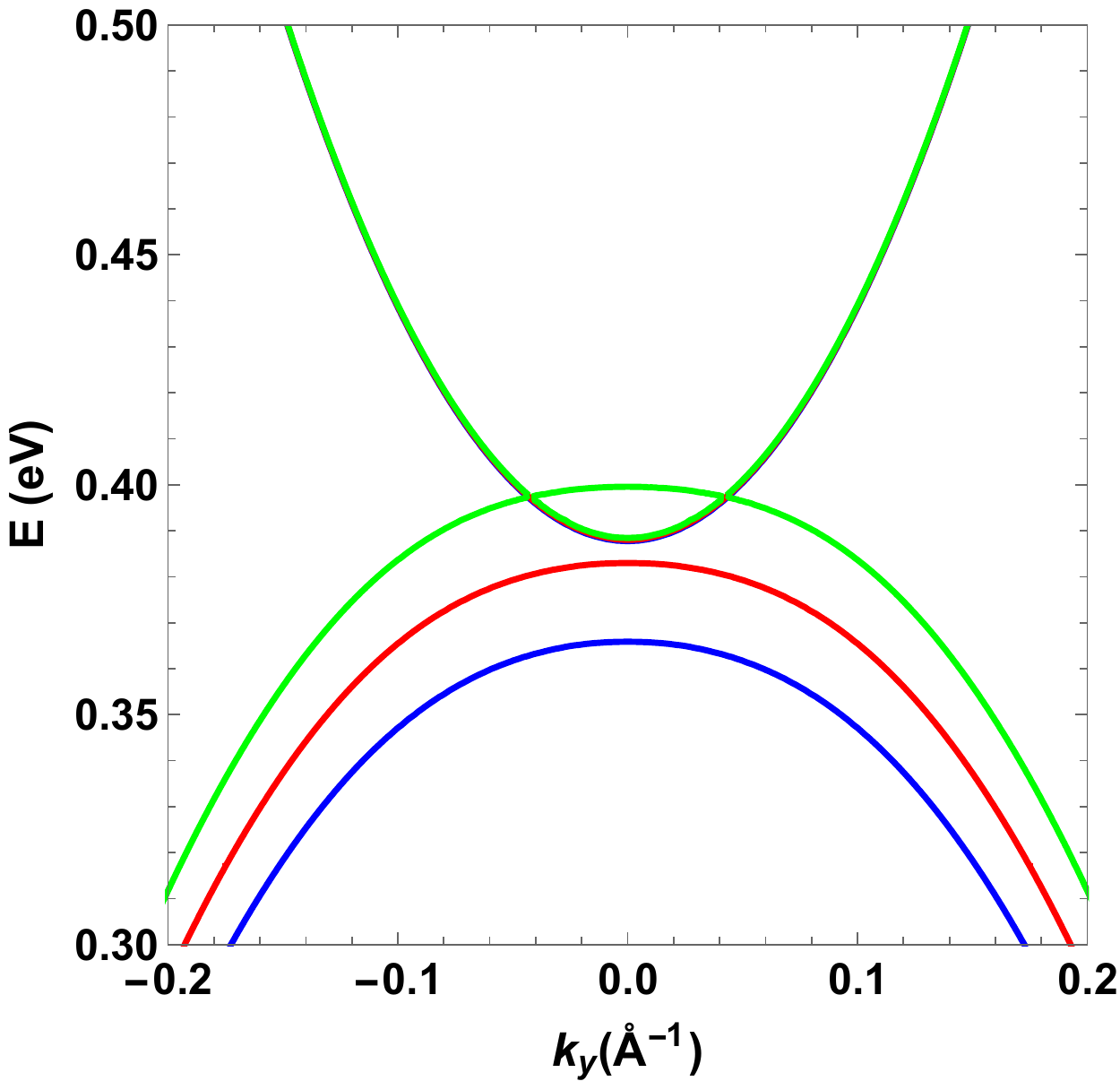}\label{f7-1}
	}\subfloat[$d=9$  nm, $d_B=2 d_W$]{
		\centering
		\includegraphics[scale=0.55]{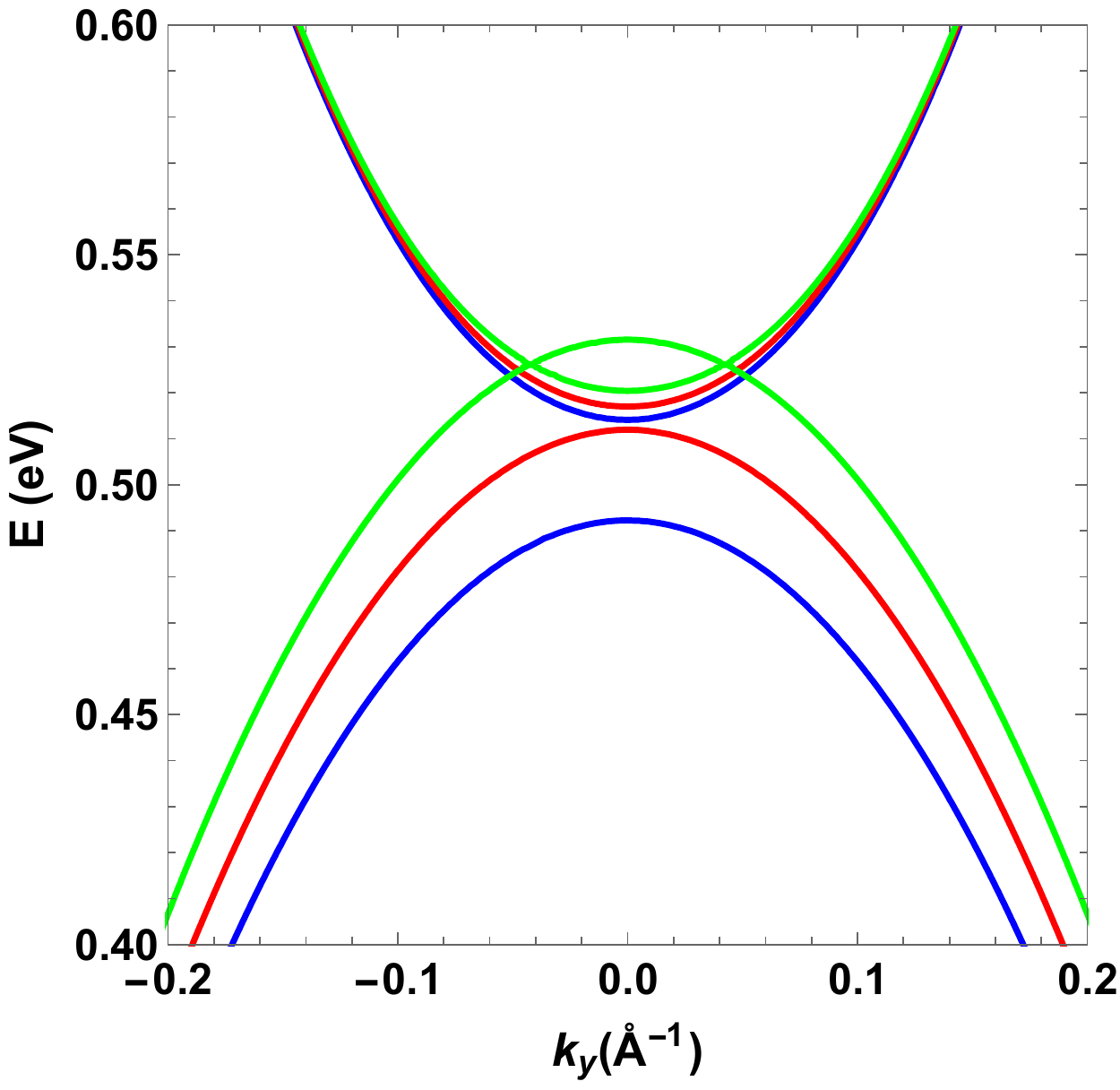}\label{f7-2}
	}

	\caption{(color online) The dispersion relation (\text{\ref{E16}}) as a function of $k_y$ for $k_x=V_W=0$, $V_B=1.72$ eV (blue), $V_B=1.74$ eV (red), $V_B=1.76$ eV (green). (a):  $d=9 $ nm, $d_W=2 d_B$ and  (b): $d=9$ nm, $d_B=2 d_W$.
	}\label{f7}
\end{figure}

We will examine the impacts of a well and barrier with different widths, i.e., a well and barrier that aren't exactly the same size. studying the electrical band structure using the dispersion relation \eqref{E16}.
 In Fig. \ref{f7}a, one observes that it is feasible to reduce and control the gap until it disappears by increasing the height barrier. This behavior is similar to that obtained for the effect of one-dimensional superlattice potentials on the
band gap of two-dimensional materials \cite{ref34}.  When $V_B$ reaches the critical value of $ V_{Bc}=1.76$ eV, we can see two Dirac points in the band structure, one on the positive side and one on the negative side, at $k_y \ne 0$. We show that the energy is symmetric for $k_y=0$. 
In Fig. \ref{f7}b for $d_B=2d_W=6$ nm, we clearly see that when the potential rises, the electron and hole energies shift up. Nevertheless,   the critical value is different and increased in comparison to the first case when we had $d_B=d_W$, but in this case,  $V_{Bc}=1.76 $ eV is the same in both cases when $d_B=2d_W$ and $d_W=2d_B$. We notice the same behavior as in the previous work with a bit of difference in the critical potential. These behaviors are similar to those obtained for the electronic structure of a graphene superlattice with massive Dirac fermions \cite{ref26}. 

\begin{figure}[H]
	\centering
	\subfloat[$d_B=3 $ nm, $d_W=2 d_B$]{
		\centering
		\includegraphics[scale=0.55]{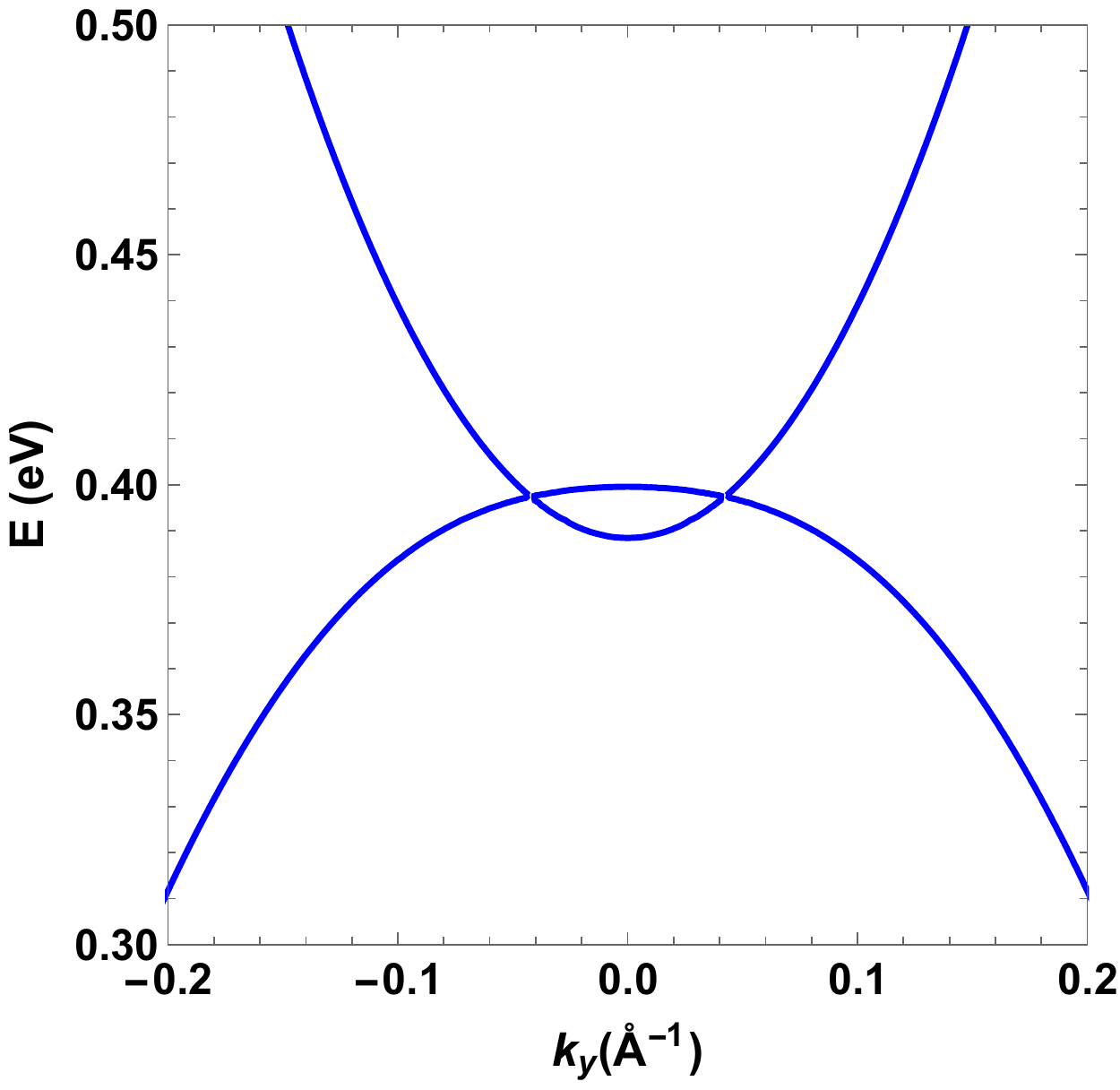}\label{f8-1}
	}\subfloat[$d_B=3$ nm, $d_W=2 d_B$]{
		\centering
		\includegraphics[scale=0.55]{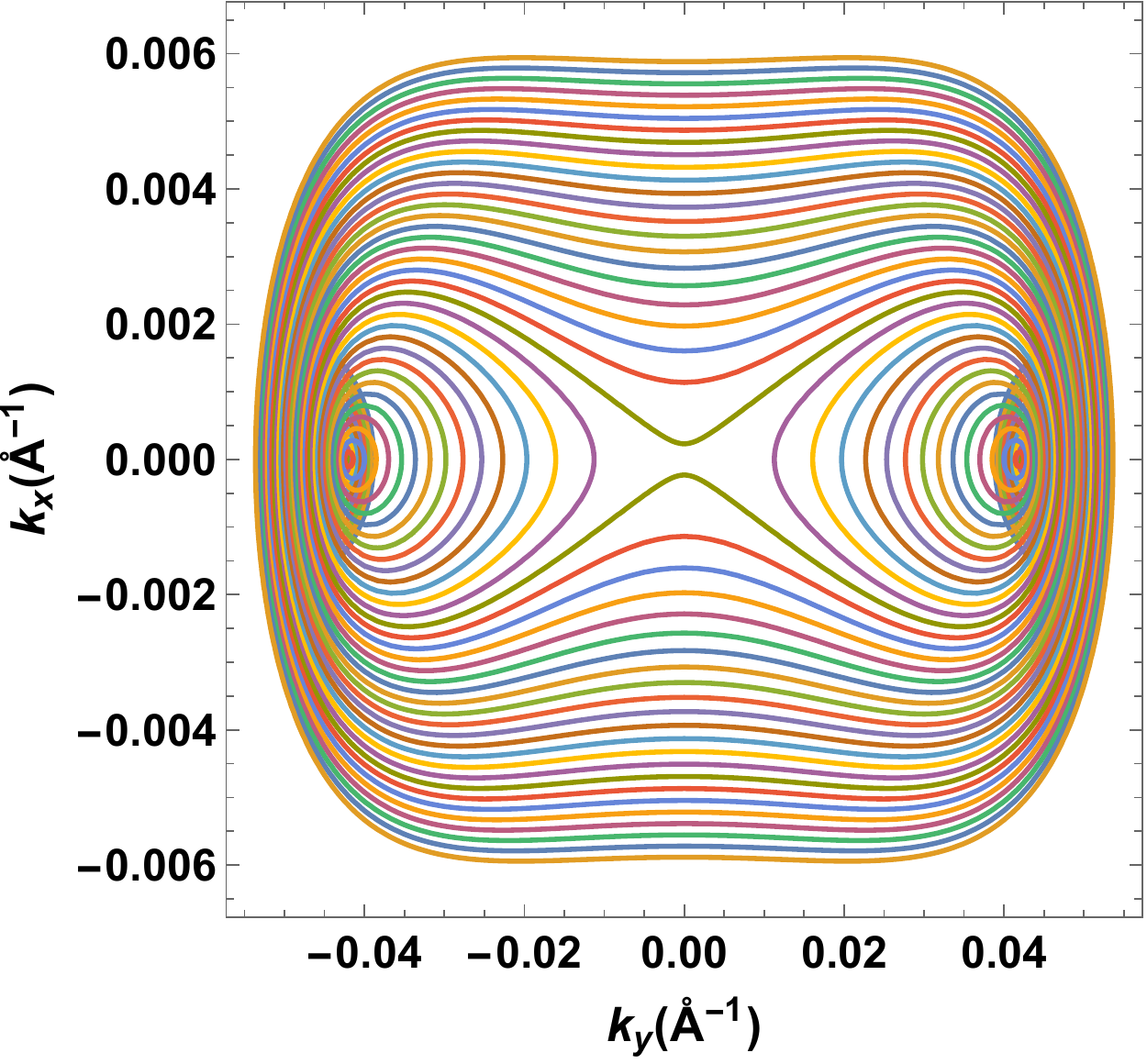}\label{f8-2}
	}	
	\caption{(color online) (a): The dispersion relation (\text{\ref{E16}}) as a function of $k_y$ for  $k_x=V_W=0$ and  $V_B=1.76$ eV. (b): Contour plot of the energy band $E \in[0.396,0.403] $ eV with step $1.5.10^{-4}$, $d=9$ nm and $V_B=1.76$ eV .
}
\label{f8}
\end{figure}
 
 The dispersion relation as a function of $ k_y $ is plotted in Fig. \ref{f8}a  for $k_x=0 $,  $d_W=2d_B=6 $ nm and  $V_B=1.76$ eV, in order to show the impact of different well and barrier widths.
 We observe two Dirac points emerge, one on the positive side and the other on the negative one. Fig. \ref{f8}b  displays the low-energy band's contour plot around the Dirac points. 
 It is clearly seen that the spectrum is symmetric about the normal incidence, there is one Dirac point on each side of the normal incidence. It is obvious that Dirac points appear when $d_W = 2d_B$. When we increase the width of the well, $d_W$, the contact points clearly shift down, as shown in \cite{ref26}. 
 
%

\begin{figure}[H]
	\centering
	\subfloat[$d_W=3 $ nm, $d_B=2 d_W$]{
		\centering
		\includegraphics[scale=0.55]{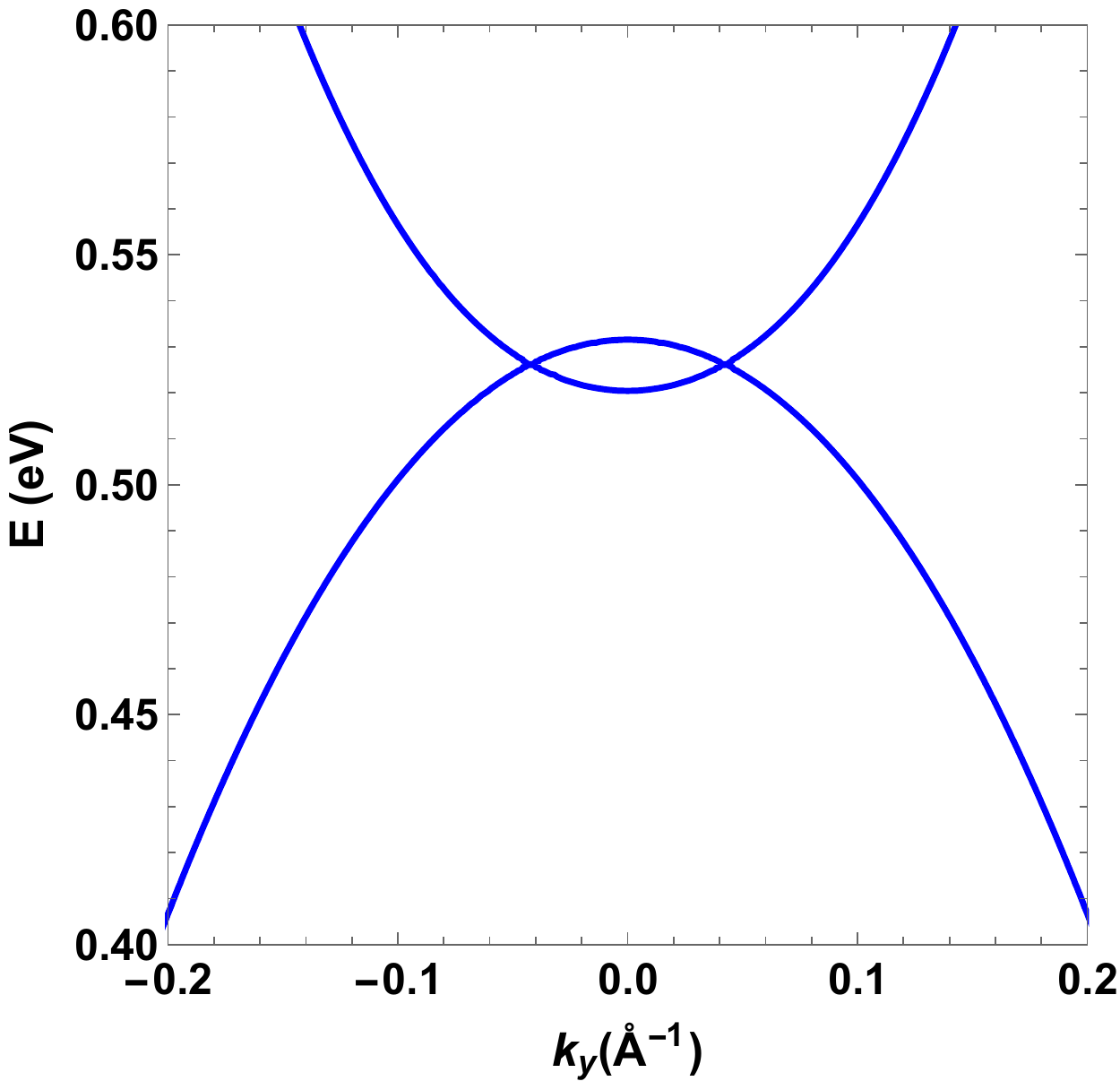}\label{f9-1}
	}\subfloat[$d_W=3 $ nm, $d_B=2 d_W$]{
		\centering
		\includegraphics[scale=0.55]{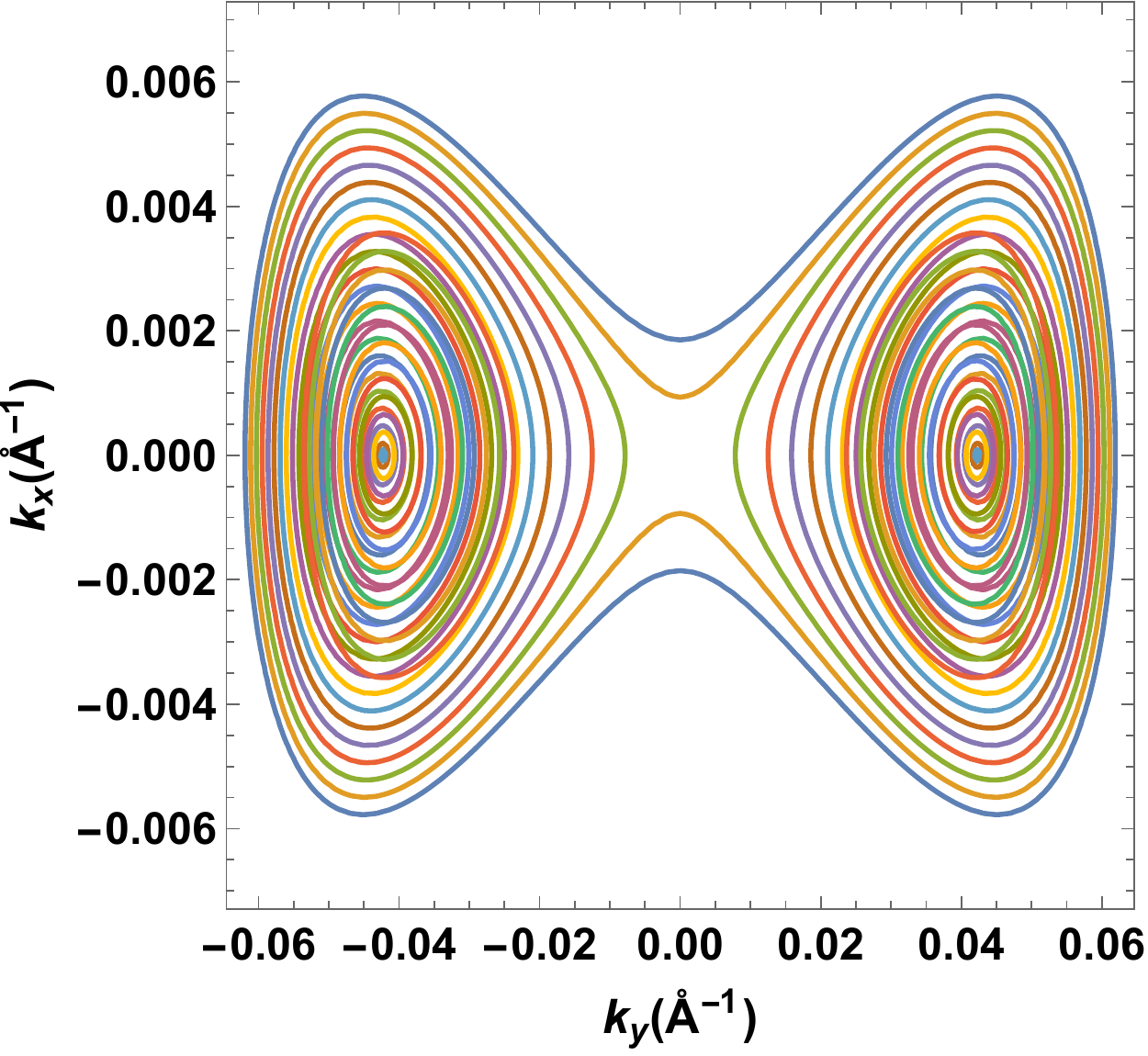}\label{f9-2}
	}
	\caption{(color online) (a): The dispersion relation (\text{\ref{E16}}) as a function of $k_y$ for $k_x=V_W=0$ and $V_B=1.76$ eV. (b): Contour plot of the energy band $E \in[0.52,0.53] $ eV with step $3.10^{-4}$, $d=9 $ nm  and $V_B=1.76$ eV.}
\label{f9}
\end{figure}

In Fig. \ref{f9}a we show  the dispersion relation as a function of $k_y $ for $k_x=V_W=0 $,  $d_B=2d_W=6 $ nm, and $V_B=1.76$ eV. We observe a pair of Dirac points  come out in the band structure, one on the positive side and the other on the negative side. Fig. \ref{f9}b demonstrates the low energy band's contour plot around the Dirac points. As can be seen, the spectrum is symmetric about the normal incidence, there is one Dirac point on each side of the normal incidence. When we increase the width barrier $d_B$, we can clearly see that the contact points have moved upward, similar to the results obtained in \cite{ref26}. 
Because we obtained similar results to those in Fig. 8 with a minor change in the placement of the Dirac points, it is clear that Dirac points rise when $d_B>d_W$ but fall when the reverse is true.

\begin{figure}[H]
	\centering
	\subfloat[$d_B=2d_W$, $d_W=100/3 $ nm]{
		\centering
		\includegraphics[scale=0.55]{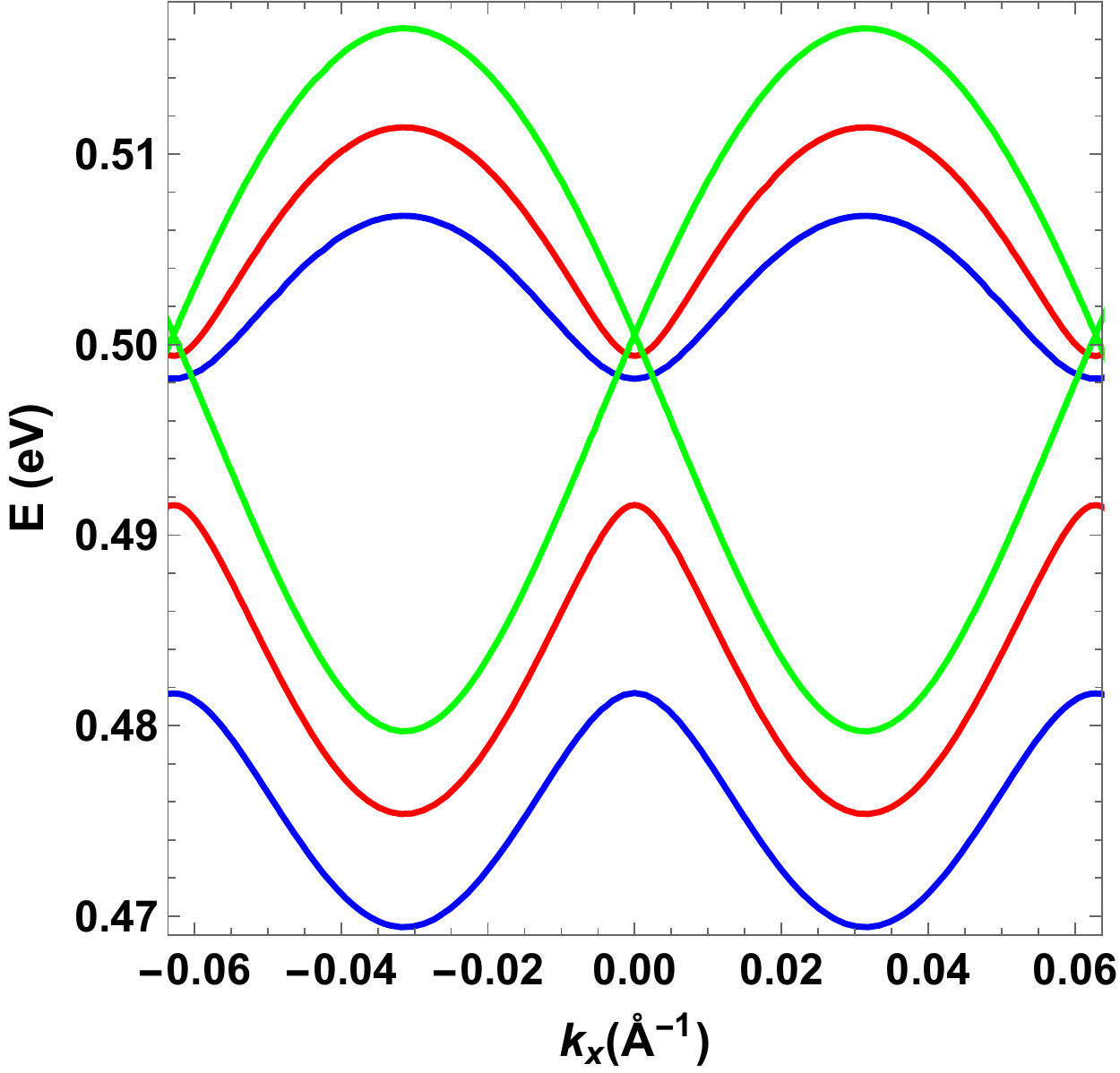}\label{f10-1}
	}\subfloat[$d_W=2d_B$, $d_B=100/3$  nm]{
		\centering
		\includegraphics[scale=0.55]{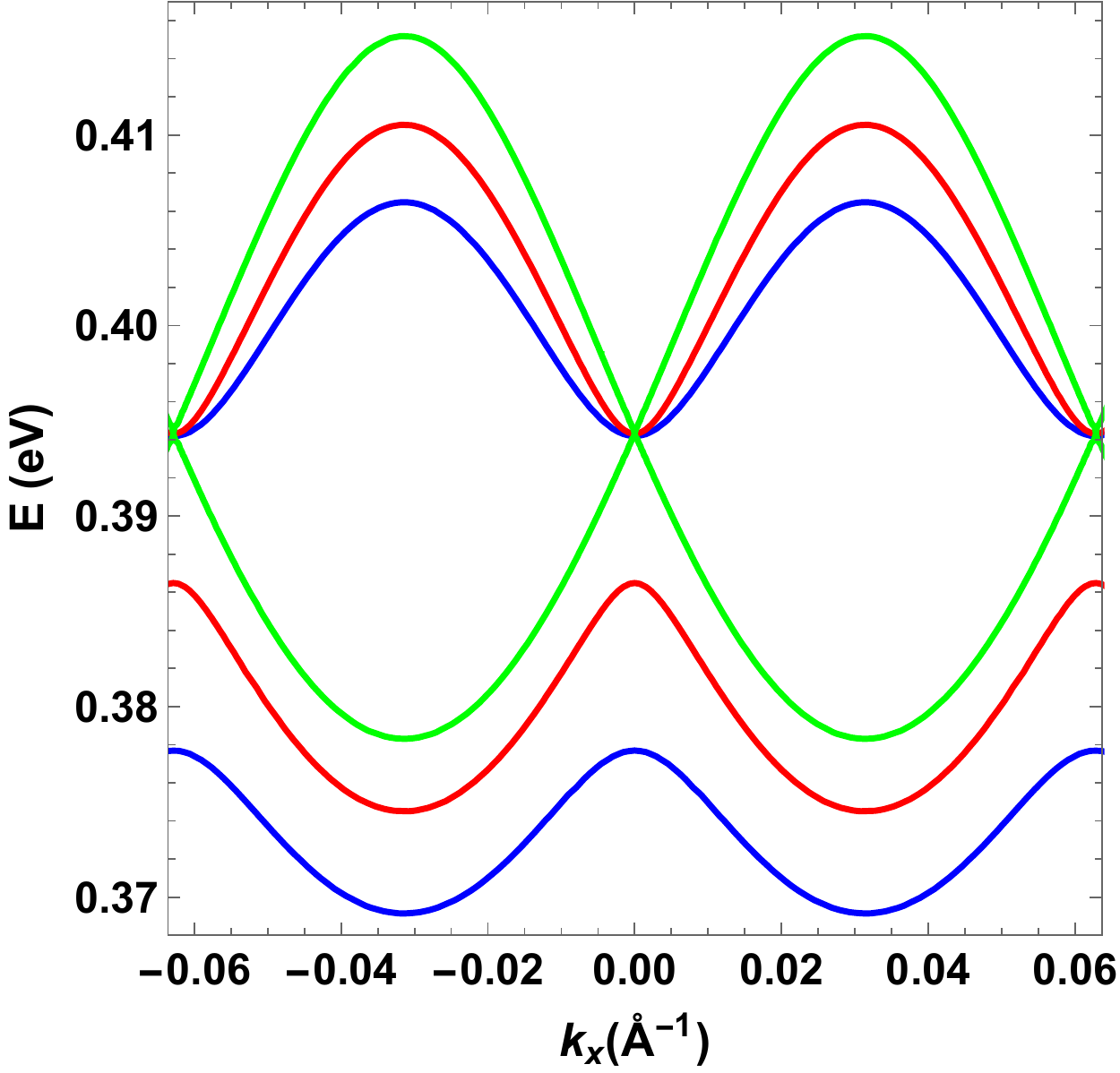}\label{f10-2}
	}
	
	\caption{(color online) The dispersion relation (\text{\ref{E16}}) as a function of $k_x$ for  $k_y=0.054$ \AA$^{-1}$, $V_W=0$, and $V_B=1.68$ eV (blue), $V_B=1.7$ eV (red), $V_B=1.706$ eV (green). (a):  $d=10 $ nm, $d_B=2 d_W$, and  (b): $d=10 $ nm, $d_W=2 d_B$.
	}\label{f10}
\end{figure}

Fig. \ref{f10} depicts the dispersion relation as a function of $k_x$ for $k_y=0.054$ \AA$^{-1}$, $V_W=0$,  $V_B=1.68$ eV (blue), $V_B=1.7$ eV (red), and $V_B=1.706$ eV (green). 
In Fig. \ref{f10}a, for $d_B=2d_W=200/3$ nm, we can tune the gap until gapless phosphorene by varying $V_B$, as happens when $V_B=1.706$ eV (green).
This clearly demonstrates that the energy can be modulated by the barrier height $V_B$.   
However, because the shift in electron energy is not the same as the shift in hole energy, there are many gaps that could all be zero for different values of $V_B$. It should be noted that the Dirac points are obtained around  $E = 0.5$  eV. 
Fig. \ref{f10}b reproduces the same results as before for $d_W=2d_B=200/3 $ nm, except that the energy is reduced by showing different shifts depending on the values of $V_B$. 
It is clear that when the potential increases, the electron and hole energies alter. The Dirac points, which are located at $E = 0.395$ eV, are slightly different in their locations, but we still see the same behavior as described in \cite{ref40}.

\section{Conclusion }\label{Secf}

Using a tight-binding effective low Hamiltonian and considering a constant periodic potential, we have examined the electrical structure and the corresponding contact points of a phosphorene superlattice. In the beginning, we derived the dispersion relation governed by an equation resulting from the boundary conditions. We were able to extract a linear behavior of the current system near the contact points and find different velocity components $(v_x, v_y)$ along the $x$- and $y$-directions by inspecting this relation. To highlight the fundamental feature of the system, we examined $(v_x, v_y) $ for the original Dirac point $(0, 0, E_0) $. As a result, we demonstrated that the barrier height and width can modulate $v_y$, exhibiting different oscillations with small amplitudes up to some values of these barrier parameters, whereas  $v_x$ decreased toward a constant.

Two scenarios were studied: one with equal well and barrier widths while maintaining a constant cell width, and the other with unequal well and barrier widths. It was demonstrated that altering the well and barrier widths or heights can be used to adjust the energy gap of the phosphorene superlattice. When the barrier height is increased, the Dirac points are moved up in two circumstances.
It was discovered that as the barrier height or the well and barrier widths increased, the energy gap shrank in both scenarios. The two situations are not identical, though. When the barrier height $V_B$ exceeds a critical value determined by the barrier width, Dirac points appear at $k_y\neq0$.Thus, the phosphorene superlattice becomes gapless under certain conditions. We demonstrated that the spectrum is symmetric in both cases when $k_y = 0$.
As a result,  to control and modify the contact points  in the phosphorene superlattice, one may adjust either the well and barrier widths or heights. 

Dirac points for fermions in phosphorene under certain constraints are very significant. In fact, the conduction and valence bands collide to generate zero-energy points, whose positions are known as Dirac points in a material's band structure. This enables effective charge transfer because electrons can readily flow between the two bands. As for the phosphorene superlattice, these Dirac points are particularly intriguing for uses in electronics and optoelectronics, as well as having potential uses in quantum computing and communication technologies. They can also be employed as waveguides for microwave or light communication systems because of their anisotropic characteristic. Additionally, they can be utilized as energy harvesters in thermoelectric devices due to their high density of states at the Fermi level. Finally, Dirac points in phosphorene offer insight into basic physics phenomena like topological insulators and Majorana fermions in addition to these possible applications.

\end{document}